\documentclass[
amsmath,amssymb,
aip,
]{revtex4-2}
\usepackage{graphicx}
\usepackage{bm}
\usepackage{color}
\usepackage{multirow}

\begin{document}


\title{A reduced variational approach for searching cycles in high-dimensional systems}

\author{Ding Wang$^1$}
\author{Yueheng Lan$^{1,2,}$}%

\email{lanyh@bupt.edu.cn}
\affiliation{%
 $^1$School of Science, Beijing University of Posts and Telecommunications, Beijing 100876, China.\\
 $^2$State Key Lab of Information Photonics and Optical Communications,Beijing University of Posts and Telecommunications, Beijing 100876, China
}%

%
%

\begin{abstract}
Searching recurrent patterns in complex systems with high-dimensional phase spaces is an important task in diverse fields.
In the current work, an improved scheme is proposed to accelerate the recently designed variational approach for finding periodic orbits in systems with chaotic dynamics based on the
existence of inertial manifold widely observed in various spatially extended systems, especially those with high dimensions.
On the premise of keeping exponential convergence of the variational method, an effective loop evolution equation is derived to greatly reduce the storage and computing time.
With repeated modification of local coordinates and evolution of the guess loop being carried out alternately, the rapid convergence and the stability of the reduction scheme are effectively achieved.
The dimension of local coordiante subspaces is generally larger than the number of nonnegative Lyapunov exponents to ensure the exponential convergence. The proposed scheme is successfully demonstrated on several well-known examples and expected to supply a powerful tool in the exploration of high-dimensional nonlinear systems.
\end{abstract}

\maketitle

\section{Introduction}\label{sec:level1}
Complex phenomena in the fields of geophysics, space physics and fluid mechanics are usually described by continuous-time dynamical systems, in which infinitely many unstable periodic orbits (cycles) play important roles in their characterization and analysis. 
 In the study of turbulence\cite{kida2001,Yamada2003}, unstable periodic orbits (UPOs) extracted from fluid flows can well describe the characteristics of turbulence, such as mean velocity profile, the time development of coherent structures, intermittency or other statistical properties. In a plethora of contexts, periodic orbits could be conveniently used to study the long-term behavior of a nonlinear system \cite{Gutzwiller1990}, for which cycle expansions are very efficient in computing average values of physical observables \cite{Cvitanovi1988,Artuso1990,LAN2010} on the strange attractors, where cycles are ordered hierarchically in terms of their topological length or stability.
Therefore, it is of great significance to develop efficient numerical methods and techniques for locating periodic orbits.

In literature, various numerical algorithms of identifying periodic orbits have been proposed so far\cite{Predrag1987,MESTEL1987,BARANGER1988,Parker1989,Stavros1995,DEANE2006,Lan2004,saiki2007},
among which the best known is probably the Newton-Raphson method.
The method and its variants\cite{MESTEL1987,BARANGER1988} show great convenience and efficiency in finding short cycles in a large variety of dynamical systems, which however for UPOs with longer periods requires good initial guesses.
Even still, it is likely to fail in finding unstable periodic orbits in high-dimensional systems\cite{kawahara2001}, because the cumulative error in the long-time evolution increases exponentially with the evolution time.
The multiple shooting algorithm is able to overcome these difficulties but a set of Poincar$\acute{\text{e}}$ sections need to be introduced to divide the whole trajectory into short segments, so as to contain the error of the long-time integration.
Nevertheless, it seems hard to select suitable Poincar$\acute{\text{e}}$ sections in high dimensions, which should intersect all the neighboring orbits transversely. To remove this trouble, the damped Newton-Raphson-Mees algorithm is proposed in Ref.\cite{saiki2007}, but the types of UPOs that can be detected by the method are limited. An alternative scheme is proposed in Ref.\cite{Lan2004}, which describes a periodic orbit with a loop of discrete points.
A variational equation derived from the original dynamics will drive the guess loop to a genuine periodic orbit exponentially. Because of the topological constraint embodied in the loop representation, irrespective of the original dynamics, the method shows excellent computational stability.

However, the variational method requires the storage and inversion of an $(Nd+1)\times(Nd+1)$ matrix ($d$ is the dimension of system and $N$ is the number of lattice points on the guess loop), which is a huge burden when exploring complex systems. 
An automatic mesh allocation scheme\cite{Zhou2008} 
and additional techniques on storage optimization are designed to alleviate these troubles to some extent in search of connecting orbits\cite{Dong2014} or periodic orbits near singularity\cite{Ding2018}. However, these strategies can not fundamentally break the bottleneck of the computational complexity in high dimensions.

In the current work, to extend application of the variational method to more stringent situation, we propose a novel scheme to greatly reduce the computation load based on the consideration below.
If a cycle is stable, any point in its neighborhood will evolve to it following the equations of motion. Therefore, for unstable cycles, only the unstable or neutral directions need to be controlled during evolution. The good news is that  in most high-dimensional systems, the number of expanding directions is much smaller than the phase space dimension. Hence, it is possible to build a numerical scheme which cares only about these unstable directions so as to accelerate the computation. The framework of the variational algorithm is especially good for this purpose. Along the guess loop, a fictitious evolution could be designed (see below) to dig out all important directions, the number of which is, say, $n$. With $n\ll d$, the complexity of the matrix computation and hence of the whole scheme is much reduced!

 The paper is organized as follows. In Sec.~\ref{sec:1}, The variational principle is briefly introduced.
In Sec.~\ref{sec:2}, The detailed derivation of the reduced variational equation and the construction and evolution of local coordinates are given.
In Sec.~\ref{sec:numerical}, three examples are used to demonstrate the validity and application details of the reduced variational approach. Our computations are summarized and discussions are made in the final Section \ref{summary}.

\section{the variational principle of cycle searching}\label{sec:1}
For a general $d$-dimensional dynamical system, its time evolution could be depicted by a set of ordinary differential equations (ODEs):
\begin{equation}\label{ODE}
 \dot{\bm{x}}(t)=\bm{v}(\bm{x}(t)),
\end{equation}
where $\bm{x} \in\mathbb{R}^d$, $t\in\mathbb{R}$ and $\bm{v}(\bm{x}): \mathbb{R}^d \to \mathbb{R}^d$ is a smooth function defined in the phase space. A periodic orbit ( or interchangeably cycle) with period $T$ is a trajectory $\bm{x}(t)$ of Eq.~(\ref{ODE}) that satisfies the condition
\begin{equation}\label{T_map}
  f^T(\bm{x})=\bm{x},
\end{equation}
where $f^T(\bm{x})$ gives the new position of $\bm{x}$ under the system dynamics Eq.~(\ref{ODE}) after a time interval $T>0$.
The variational method and its variants have been widely used to locate cycles in many systems\cite{Lan2004,Ding2018,Dong2014} and show impressive robustness and convergence,
which deploy a loop in the phase space being driven to a true cycle by the fictitious evolution equation
\begin{equation}\label{vari}
  \frac{\partial^2\tilde{\bm{x}}}{\partial s\partial \tau}-\lambda\frac{\partial\bm v}{\partial\bm x}\frac{\partial\tilde{\bm{x}}}{\partial \tau}-\bm v\frac{\partial\lambda}{\partial\tau}=\lambda\bm v -\tilde{\bm v},
\end{equation}
where $\tilde{\bm{x}}$ marks the representative points of the loop and is parameterized by $s~(s\in[0,2\pi])$, $\tilde{\bm v}=\frac{\partial\tilde{\bm x}}{\partial s}~(\tilde{\bm{x}}\in\mathbb{R}^d)$ is the loop velocity, $\bm v$ is the velocity field of the dynamical system given by Eq.~(\ref{ODE}) and $\tau$ is the fictitious time which records the evolution from an initial guess loop to the desired periodic orbit. The parameter $\lambda$ is independent of $s$ but is a function of $\tau$ which is used to match the magnitude of $\bm v$ and $\tilde{\bm v}$, or in other words, to adjust the period. We may rewrite Eq.~(\ref{vari}) as
\begin{equation}\label{Eq1}
  \frac{\partial}{\partial\tau}(\tilde{\bm v}-\lambda\bm v)=\lambda\bm v-\tilde{\bm v},
\end{equation}
the solution of which could be formally written as
\begin{equation}\label{Eq_s}
  \tilde{\bm v}-\lambda\bm{v}=e^{-\tau}(\tilde{\bm{v}}-\lambda\bm{v})|_{\tau=0},
\end{equation}
showing that the error $\tilde{\bm v}-\lambda\bm{v}$ exponentially decreases with $\tau$ and the loop exponentially approaches the periodic orbit.

In a discretization of the loop, the vector $\tilde{\bm v}$ is calculated by a five-point approximation scheme for numerical stability:
\begin{align}\label{tiliderv}
  \tilde{\bm v}(s_i)&\equiv \frac{1}{12h}[8\tilde{\bm x}(s_{i+1})-8\tilde{\bm x}(s_{i-1})+\tilde{\bm x}(s_{i-2})-\tilde{\bm x}(s_{i+2})]\nonumber\\
  &=\hat{\bm D}\tilde{\bm x}(s_i)
\end{align}
in which $\hat{\bm{D}}$ is a matrix operator with the periodic condition
\begin{equation}\label{D}
\hat{\bm D}=\frac{1}{12h}
  \begin{pmatrix}
       ~~0 &~~8 & -1   &  ~    &  ~   &  ~   &~~1    & -8 \\
        -8 &~~0 &~~8   & -1    &  ~   &  ~   &  ~    &~~1 \\
       ~~1 & -8 &~~0   &~~8    & -1   &  ~   &  ~    &  ~ \\
         ~ &~~1 & -8   &~~0    &~~8   & -1   &  ~    &  ~ \\
         ~ &  ~ &\ddots&\ddots &\ddots&\ddots& \ddots&  ~ \\
         ~ &  ~ &  ~   &~~1    & -8   &~~0   &~~8    & -1 \\
        -1 &  ~ &  ~   &  ~    &~~1   & -8   &~~0    &~~8 \\
       ~~8 & -1 &  ~   &  ~    &  ~   &~~1   & -8    &~~0
  \end{pmatrix}
\end{equation}
where every element 
is a $d\times d$ diagonal matrix and $h=2\pi/N$, $N$ being the number of lattice points. The discretized version of Eq.~(\ref{vari}) with a fictitious time interval $\delta\tau$ is
\begin{equation}\label{v_Euler}
  \begin{pmatrix}
       \hat{\bm{A}} &~-\hat{\bm{v}} \\
       \hat{\bm{a}} &~~~0
  \end{pmatrix}
    \begin{pmatrix}
       \delta\hat{\bm{x}} \\
       \delta\lambda
  \end{pmatrix}=\delta\tau
  \begin{pmatrix}
       \lambda\hat{\bm{v}}-\hat{\tilde{\bm{v}}}\\
       0
  \end{pmatrix}
\end{equation}
with  $$\hat{\bm{A}}=\hat{\bm D}-\lambda \text{diag}[A_1,~A_2,...,A_N]$$
where $A_n$ is the velocity gradient matrix $A_n=\frac{\partial v_i}{\partial v_j}|_{s=s_n}$, $\hat{\bm{v}}=(\bm{v}_1,\bm{v}_2,...,\bm{v}_n)^t$ and $\hat{\tilde{\bm{v}}}=(\tilde{\bm{v}}_1,\tilde{\bm{v}}_2,...,\tilde{\bm{v}}_n)^t$ are the velocity vector of the original flow and the loop velocity, respectively. $\hat{\bm{a}}$ is an $Nd$-dimensional row vector which imposes a gauge-fixing condition on the coordinate corrections $ \delta\hat{\bm{x}}$ in order to remove the neutral direction along which the cycle is invariant upon shifting all the points.
In each iteration, we set up and numerically solve Eq.~(\ref{v_Euler}) and update the coordinates of the guess loop which is supposed to approach exponentially a periodic orbit if it exists.

\section{A reduced numerical Scheme }\label{sec:2}
Most computation load originates from the matrix $\tiny{\begin{pmatrix}\hat{\bm{A}} &~-\hat{\bm{v}} \\ \hat{\bm{a}} &~~~0\end{pmatrix}}$ in Eq.~(\ref{v_Euler}) which is $(Nd+1)\times (Nd+1)$, involving the dimension $d$ of the phase space and the number of lattice points $N$. With the increase of $d$ and $N$, the required storage and computing time quickly increases, which is a major bottleneck preventing a wide application of the variational method in high-dimensional systems. In Ref.~\cite{Ding2018}, an automatic allocation scheme of lattice points is adopted to minimize $N$. However, in a complex system, the dimension $d$ of the system is much greater than $N$. Only reducing the number of lattice points is far from enough, and thus we try to accelerate the variational method by reducing the effective  dimension of the local coordinate system.
\subsection{Reduction by projection}\label{sec:2_1}
In practice, not all directions are equally important for orbit adjustment when approaching a periodic orbit from a guess loop. On a hyperplane perpendicular to the  periodic orbit, the vicinity of an orbit point is stretched in a few directions and compressed in others when moving along the orbit.
The correction in the compressed directions is automatically made during the orbit evolution and the deviation in the stretching or neutral directions has to be corrected by the variational scheme. Therefore, it is essential to find these important directions and rewrite the original variational equation in reduced coordinate frames.

\begin{figure}[]
  \center\includegraphics[width=8cm]{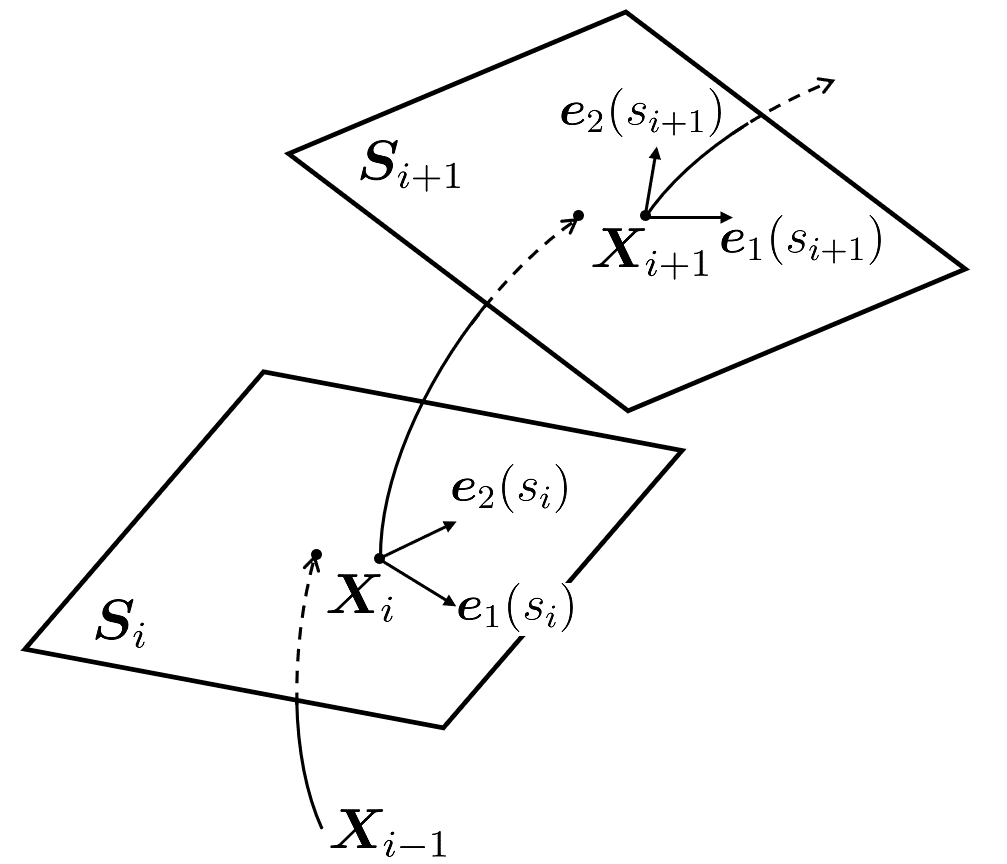}
    \caption{\label{Fig1} The local coordinate frames $\{\bm{e}_k(s_i)\}$ $(k=1,2,...,n;~i=1,2,...,N$ and $n\leq d)$. along the orbit parametrized by $s_i$, where $S_i$ represents the hyperplane spanned by the vectors $\bm{e_{k}}(s_i)~(k\le d)$.
    }
\end{figure}

Assuming that along the important directions, we already built a family of local coordinate frames $\{\bm{e}_k(s_i)\}$ $(k=1,2,...,n;~i=1,2,...,N$ and $n\leq d)$ for the projection ( Fig.\ref{Fig1}), the reduced variational equation may be derived as follows.

Firstly, we write an approximation of Eq.(\ref{tiliderv}) in the subspace $\{\bm{e}_k\}$,
\begin{small}
\begin{align}\label{tiliderv2}
  \tilde{v}_j(s_i)&=\frac{1}{12h}[8\tilde{ x}_j(s_{i+1})\bm{e}_j(s_{i+1})-8\tilde{ x}_j(s_{i-1})\bm{e}_j(s_{i-1})
  +\tilde{ x}_j(s_{i-2})\bm{e}_j(s_{i-2})-\tilde{x}_j(s_{i+2})\bm{e}_j(s_{i+2})],
\end{align}
\end{small}
and a dot-product of both sides of Eq.(\ref{vari}) with $\bm{e}_k(s_i)$ results in
\begin{equation}\label{vari2}
   \bm{e}_k(s_i)\cdot\bigg[\frac{\partial^2\tilde{\bm{x}}}{\partial s\partial \tau}-\lambda\frac{\partial\bm v}{\partial\bm x}\frac{\partial\tilde{\bm{x}}}{\partial \tau}-\bm v\frac{\partial\lambda}{\partial\tau}\bigg]=\bm{e}_k(s_i)\cdot\bigg[\lambda\bm v -\tilde{\bm v}\bigg].
\end{equation}
By feeding Eq.~(\ref{tiliderv2}) into Eq.~(\ref{vari2}), we obtain the projection of the first term in Eq.(\ref{vari2}),
\begin{small}
\begin{equation}\label{Eq2}
  \bm{e}_k(s_i)\cdot\Delta\tilde{\bm v}(s_i)=\frac{1}{12h}
  \bigg[\Delta\tilde{x}_{i-2}^j\cdot(\bm{e}^{k}_i\cdot\bm{e}^{j}_{i-2})-8\Delta\tilde{ x}_{i-1}^j(\bm{e}^{k}_i\cdot\bm{e}^{j}_{i-1})
  +8\Delta\tilde{x}_{i+1}^j\cdot(\bm{e}^{k}_i\cdot\bm{e}^{j}_{i+1})-\Delta\tilde{ x}_{i+2}^j\cdot(\bm{e}^{k}_i\cdot\bm{e}^{j}_{i+2})\bigg],
\end{equation}
\end{small}
where $\Delta\tilde{x}^j_{i}$ and $\bm{e}^{j}_{i}$ labels $\Delta\tilde{ x}_j(s_{i})$ and $\bm{e}_j(s_{i})$ respectively. According to Eq.~(\ref{Eq2}), every constant block in the differential operator $\hat{\bm D}$ in Eq.~(\ref{D}) is replaced by an $n\times n$ matrix which is defined by $\bm{e}^{k}_i\cdot\bm{e}^{j}_{i'}$, where $(k,j)$ marks the block position and $(i,i^{\prime})$ the position within the $(k,j)-$block. The second term $\frac{\partial\bm v}{\partial\bm x}\frac{\partial\tilde{\bm{x}}}{\partial \tau}$ in Eq.(\ref{vari2}) can be written as
\begin{align}\label{Eq3}
  \bm{e}_k(s_i)\cdot\Delta\bm v(s_i)
         &=\bm{e}_k(s_i)\cdot\bigg[\bm v\bigg(\tilde{\bm x}_i+\sum\limits_j\Delta x_{ij}\bm{e}_j\small(s_i\small)\bigg)-\bm v(\tilde{\bm x}_i)\bigg]\nonumber\\
         &=\bm{e}_k(s_i)\cdot\sum\limits_p\bigg[\hat{\bm e}_p v_p\big(\tilde{\bm x}_i+\sum\limits_j\Delta x_{ij}\bm{e}_j\small(s_i\small)\big)-\hat{\bm e}_p  v_p(\tilde{\bm x}_i)\bigg]\nonumber\\
         &=\bm{e}_k(s_i)\cdot\sum\limits_p\bigg[\hat{\bm e}_p v_p(\tilde{\bm x}_i)+
        \hat{\bm e}_p\sum\limits_j\big(\frac{\partial v_p}{\partial x_j} \Delta x_{ij}\big) -\hat{\bm e}_p  v_p(\tilde{\bm x}_i)\bigg]\nonumber\\
         &=\sum\limits_p\bigg[
        {e}_{kp}(s_i)\sum\limits_j\big(\frac{\partial v_p}{\partial x_j} \Delta x_{ij}\big) \bigg]\nonumber\\
         &=\sum\limits_{j}\frac{\partial v_k}{\partial x_j}\Delta x_{ij}
\end{align}
where $k\,,j$ marks the new coordinates and $\{\hat{\bm e}_p\}$ make up the original Cartesian coordinate system which are constant at every point in the phase space. $ v_k=\bm{\bm v}\cdot\bm{e}_k(s_i)$ ($\{\bm{e}_k\}$ is an orthogonal basis), $e_{kp}(s_i)=\bm{e}_k(s_i)\cdot\hat{\bm{e}}_p$, 
 and $\Delta\tilde{\bm x}(s_i)=\sum\limits_j\Delta x_{ij}\bm{e}_j(s_i)=\Delta\tilde{\bm x}_i$. Therefore, the size of the velocity gradient matrix $\frac{\partial\bm v_k}{\partial\bm x_j}$, which is obtained by numerical differentiation, is much smaller than that of the original velocity gradient $A=\partial v/\partial x$ of Eq.(\ref{vari}) if $n\ll d$.
About other terms in Eq.(\ref{vari2}), the vectors can be just projected into the local subspace $\{\bm{e}_k\}$ and the matrix $\tiny{\begin{pmatrix}\vec{\bm{A}} &~-\vec{\bm{v}} \\ \vec{\bm{a}} &~~~0\end{pmatrix}}$  in Eq.~(\ref{v_Euler}) is then reduced to $(Nn+1)\times (Nn+1)$.
Because $n \ll d$ actually holds in many nonlinear systems, the reduced matrix order $Nn+1$ will be much smaller than the original one $Nd+1$.
Therefore the storage and computing time required to deal with the reduced form of Eq.~(\ref{v_Euler}) are greatly cut down. The larger the dimension $d$, the more prominent the benefits of this reduction.
\subsection{Construction and evolution of the local coordinates}\label{sec:2}
In this section, we will discuss how to build the local coordinate system $\{\bm{e}_k\}$ mentioned above at each lattice point $s_i$. However,
before that, Let's first explain the concept of  pseudo-evolution.
Two adjacent points on the loop are approximately related by the evolution $\bm x(s_i)\longrightarrow\bm x(s_{i+1})$ with Eq.~(\ref{ODE}) if the guess is reasonably good. Similarly, the corresponding nearby points are assumed evolving as $\bm x(s_i)+\bm h(s_i)\longrightarrow\bm x(s_{i+1})+\bm h(s_{i+1})$ governed by Eq.(\ref{ODE}) either, where $\bm h(s_i)$ and $\bm h(s_{i+1})$ represent respectively tiny offset vectors at $\bm{x}(s_i)$ and $\bm{x}(s_{i+1})$. Then we have approximately
\begin{equation}\label{h}
  \bm h(s_{i+1})\longrightarrow \bm h(s_i)+[\bm v(\bm x_i+\bm h(s_i))-\bm v(\bm x_i)]\Delta t,
\end{equation}
where $\Delta t$ is the time interval between points $s_i$ and $s_{i+1}$. Eq.~(\ref{h}) is the pseudo-evolution equation of points near the guess loop, which obviously is rough when the guess loop is far away from the periodic orbit we are looking for. As the loop gradually approaches the periodic orbit, it becomes more and more accurate.

\begin{figure}[]
 \center\includegraphics[width=12cm]{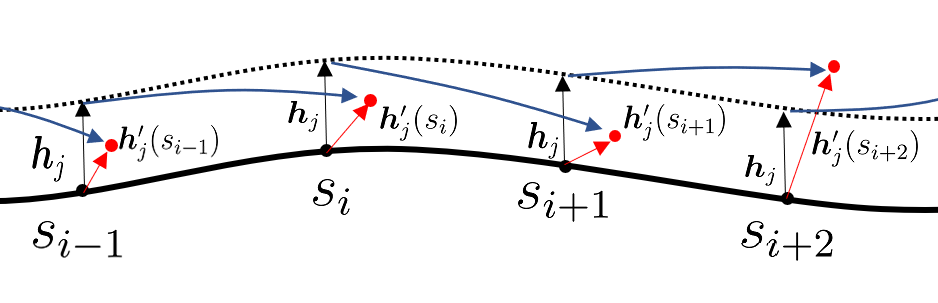}
\caption{\label{Fig3}
The evolution of local coordinate frame. The thick black solid line represents the guess periodic orbit $\bm{X}(s_i)$, and the dotted line is an adjacent curve $\bm{X}_{near}(s_i)=\bm{X}(s_i)+\bm{h}_{j}$ near the periodic orbit, and $|\bm{h}_{j}| = 10^{-3}$. $s_i$ is the curvilinear coordinate of the point on the guess orbit. $\bm{h}^{\prime}_j(s_{i+1})$ (red arrow) at $s_{i+1}$ evolves from $\bm{h}_j$ (black arrow) at $s_i$.}
\end{figure}

In order to decently depict the evolution along the orbit, we employ a second-order numerical scheme based on Eq.~(\ref{h}),
\begin{align}\label{h2}
&\bm v_1=[\bm v(\bm x_i)+\bm v(\bm x_{i+1})]/2, \nonumber\\
&\bm v_2=[\bm v(\bm x_i+\bm h_j(s_i))+\bm v(\bm x_{i+1}+\bm h_j(s_{i+1}))]/2,\\
&\bm h'_j(s_{i+1})\longrightarrow \bm h_j(s_i)+(\bm v_2-\bm v_1)\Delta t,\nonumber
\end{align}
where $\bm h_j(s_i)$ is a tiny vector in the $j$-$th$ direction $\bm e_j(s_i)$, $\bm h'_j(s_{i+1})$ is the tiny vector at $s_{i+1}$ after one step of evolution, while $\bm h_j(s_{i+1})$ is the vector at point $s_{i+1}$ before the evolution. In order to ensure that the new coordinate system is orthogonal, after each evolution it is necessary to carry out  the Schmidt orthogonalization and normalization of $\{\bm h_j^{\prime}(s_{i})\}$ to get the new set $\{\bm e_j(s_{i})\}$ at each $s_{i}$,
\begin{align}\label{hh3}
&\bm h_j^{\prime}(s_{i})\longrightarrow \bm h_j^{\prime}(s_i)-\sum_{k=1}^{j-1}c_k\bm e_k(s_i),\nonumber\\
&\bm e_j(s_{i})\longrightarrow \bm h_j^{\prime}(s_i)/|\bm h_j^{\prime}(s_i)|,\nonumber
\end{align}
where $c_k=\bm e_k(s_i)\cdot\bm h^{\prime}_j(s_i)$.
As shown in Fig.~\ref{Fig3},  the evolution of the local coordinate axis in the $j$-$th$ direction is plotted.
For each evolution, a very small vector $\bm{h}_j$ (black arrow) in some $j$-th direction at each lattice point $s_i$ evolves to $\bm{h}_j^{\prime}(s_{i+1})$ (red arrow) at $s_{i+1}$. After orthomalizing the vector $\bm{h}_j^{\prime}(s_{i+1})$, we get a new direction vector $\bm{e}_j(s_{i+1})$ (not displayed) at $s_{i+1}$.
Repeating the procedure in all involved directions we will finally build up a new coordinate frame at $s_{i+1}$.

What is a proper dimension of the local coordinate system? Obviously, according to previous arguments it should not be less than the number of positive Lyapunov exponents of the system.
The Lyapunov exponent equals $0$ in the direction of the velocity field, but the sampling points $x(s_i)$ will approach the desired periodic orbit only along stable directions.
Therefore, we include the velocity vector in the reduced coordinate subspace, as well as $2\sim 3$ other not-so-unstable or neutral directions to accelerate the convergence.
To start with, at each lattice point we may select the same first few coordinate directions in the full phase space to build the initial local frame.
Obviously, such a coordinate system is not in the stretching direction of the orbit, and we should have evolved it for some time to achieve that. Nevertheless, it does not hinder us from computing corrections in this frame and modifying it along the way.
As the guess loop gradually approaches the periodic orbit by repeatedly solving Eq.~(\ref{v_Euler})) 
, modifications of the coordinate system are made with the scheme illustrated in Fig.~\ref{Fig3}.
In each modification, each local coordinate frame moves along the loop for $n_1$ steps (roughly $5\sim 15$ steps in the current work). If the total number of modifications during the whole search is $n_2$, in order to largely cover the expanding directions, $n_1n_2\Delta t \ge 1/\lambda _L$ should be satisfied, where $\lambda _L$ is some kind of average Lyapunov exponent. In the following, we find that $n_1\times n_2 > N/2$ could do the job quite well, where $N$ is the number of lattice points depicting the periodic orbit.

We start the cycle search by numerical integration of the dynamical system under investigation and  locate possible orbit segments that nearly repeat themselves - the recurring segments. Numerical experiments reveal regions where a trajectory spends most of its life, giving us the first hunch where to initialize a loop. After an extensive search for the recurring orbits, to start the variational scheme, we first take a spatial FFT of one such segment and keep only the lowest wavenumber components, of which an inverse Fourier transform back to the phase space yields a smooth loop , which will be used as our initial guess. In the following, we will apply the reduced scheme of the variational method discussed above to several examples.

\section{Application and discussion}\label{sec:numerical}
In this section, we will use some examples to show the validity of the reduced variational method. The convergence criterion in the following is
\begin{equation}\label{F}
  Err.=\frac{1}{N}\sqrt{\sum\limits_{i=1}^{N}(\tilde{v}_i-\lambda v_i)^2}\leq10^{-6}.
\end{equation}

\subsection{Navier-Stokes equation with periodic boundary conditions}
The incompressible Navier-stokes equation on the torus $T^2=[0,2\pi]\times[0,2\pi]$ is written as\cite{Boldrighini1979,Franceschini1981}
\begin{align}\label{N_S}
  \frac{\partial\bm u}{\partial t}+(\bm u\cdot\nabla)\bm u =& -\nabla p + f +\nu\Delta\bm u,\\
  div~\bm u=&0,\\
  \int_{T^2}\bm u~dx =&0,
\end{align}
where $\bm u$ and $p$ is the velocity field and the pressure respectively, and $f$ is the external periodic driving. Under certain conditions, Eq.~(\ref{N_S}) could be expanded with $9$ Fourier modes\cite{2014nine-mode}, which reads
\begin{align}\label{ODE91}
  \dot{x_1} =& -2x_1+4x_3x_5+4x_2x_4-4x_6x_9-4x_7x_8, \nonumber\\
  \dot{x_2} =& -9x_2+3x_1x_4, \nonumber\\
  \dot{x_3} =& -5x_3-x_1x_5 ,\nonumber\\
  \dot{x_4} =& -5x_4-7x_1x_2+\frac{9\sqrt{5}}{5}x_1x_7+Re, \nonumber\\
  \dot{x_5} =& -x_5-\sqrt{5}x_1x_6-3x_1x_3, \\
  \dot{x_6} =& -x_6+3x_1x_9+\sqrt{5}x_1x_5,\nonumber\\
  \dot{x_7} =& -5x_7-\frac{9\sqrt{5}}{5}x_1x_4+7x_1x_8,\nonumber\\
  \dot{x_8} =& -9x_8-3x_1x_7, \nonumber\\
  \dot{x_9} =& -5x_9+x_1x_6,\nonumber
\end{align}
where $R_e= 58$ is the Reynolds number.
 \begin{figure}[]
  \includegraphics[width=7cm]{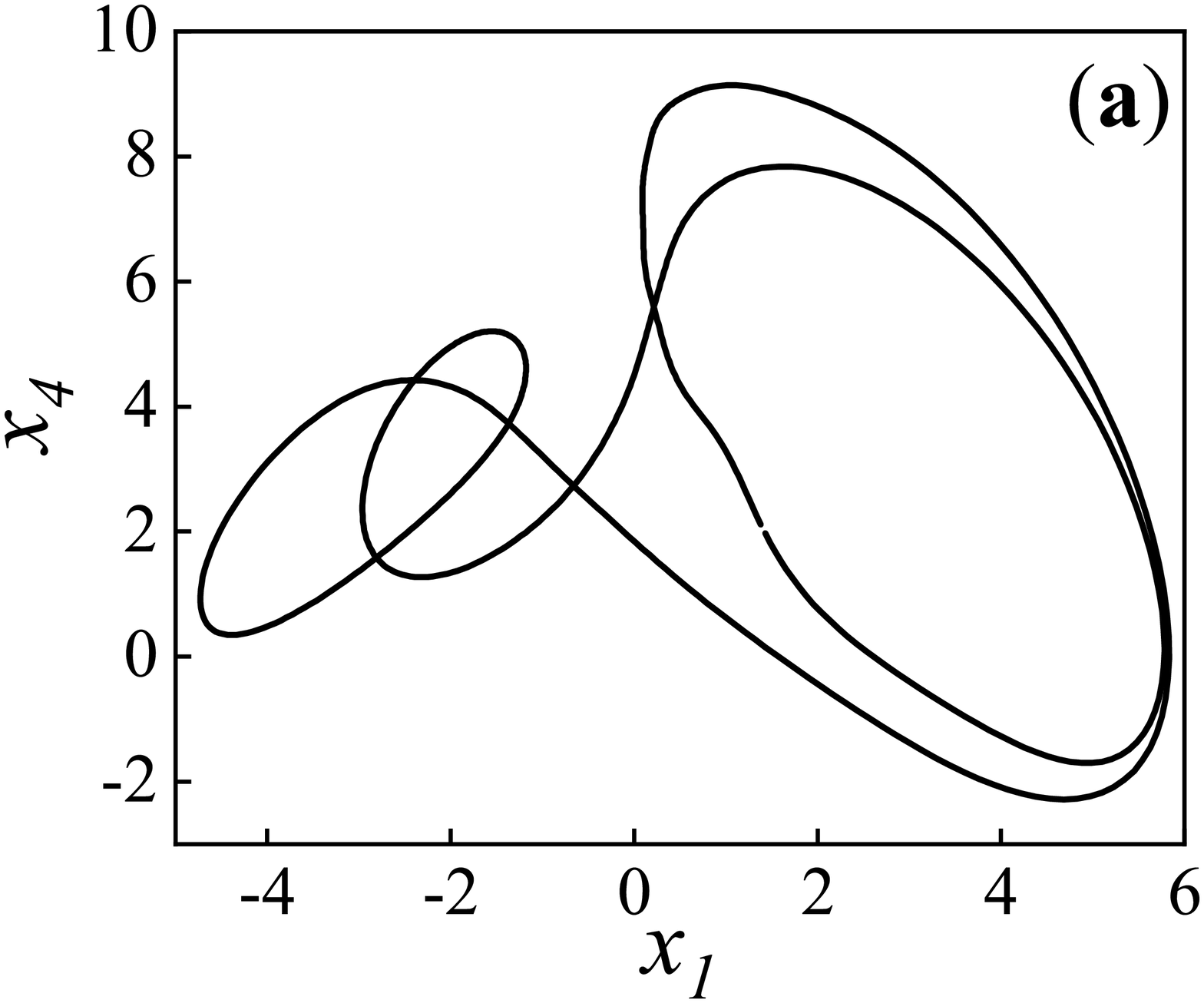}
  \includegraphics[width=7cm,height=5.55cm]{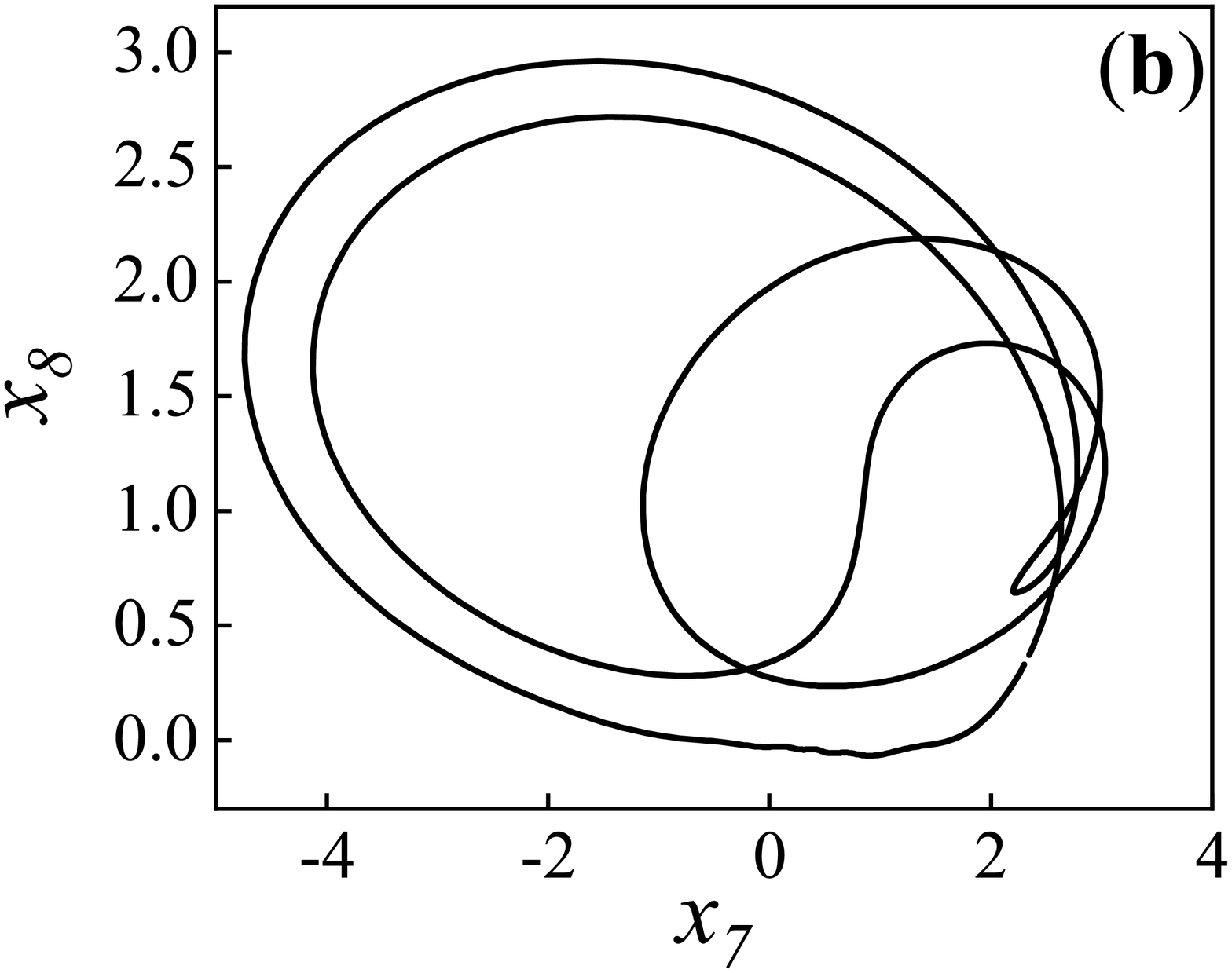}\\
  \includegraphics[width=7cm]{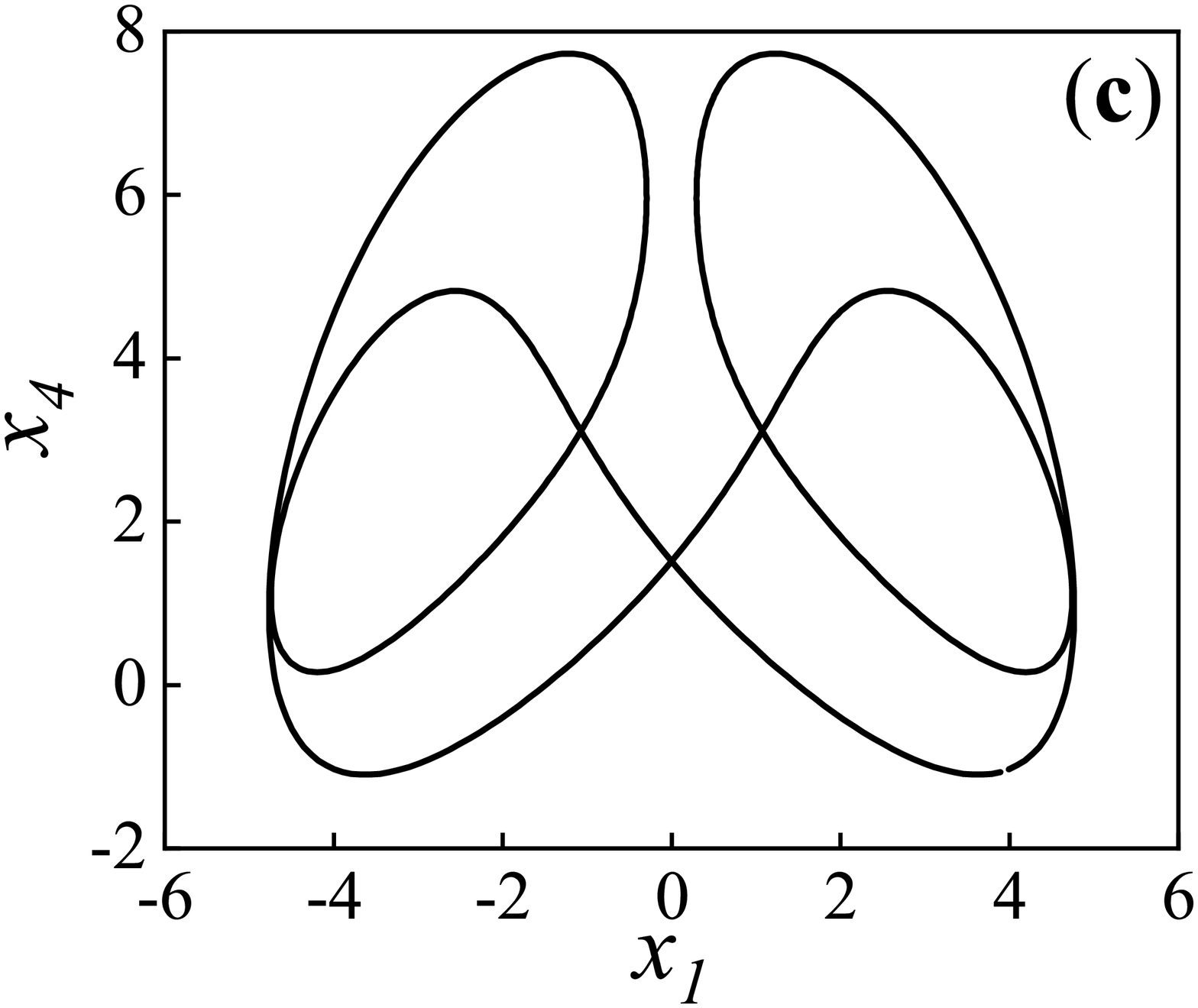}
  \includegraphics[width=7cm,height=5.85cm]{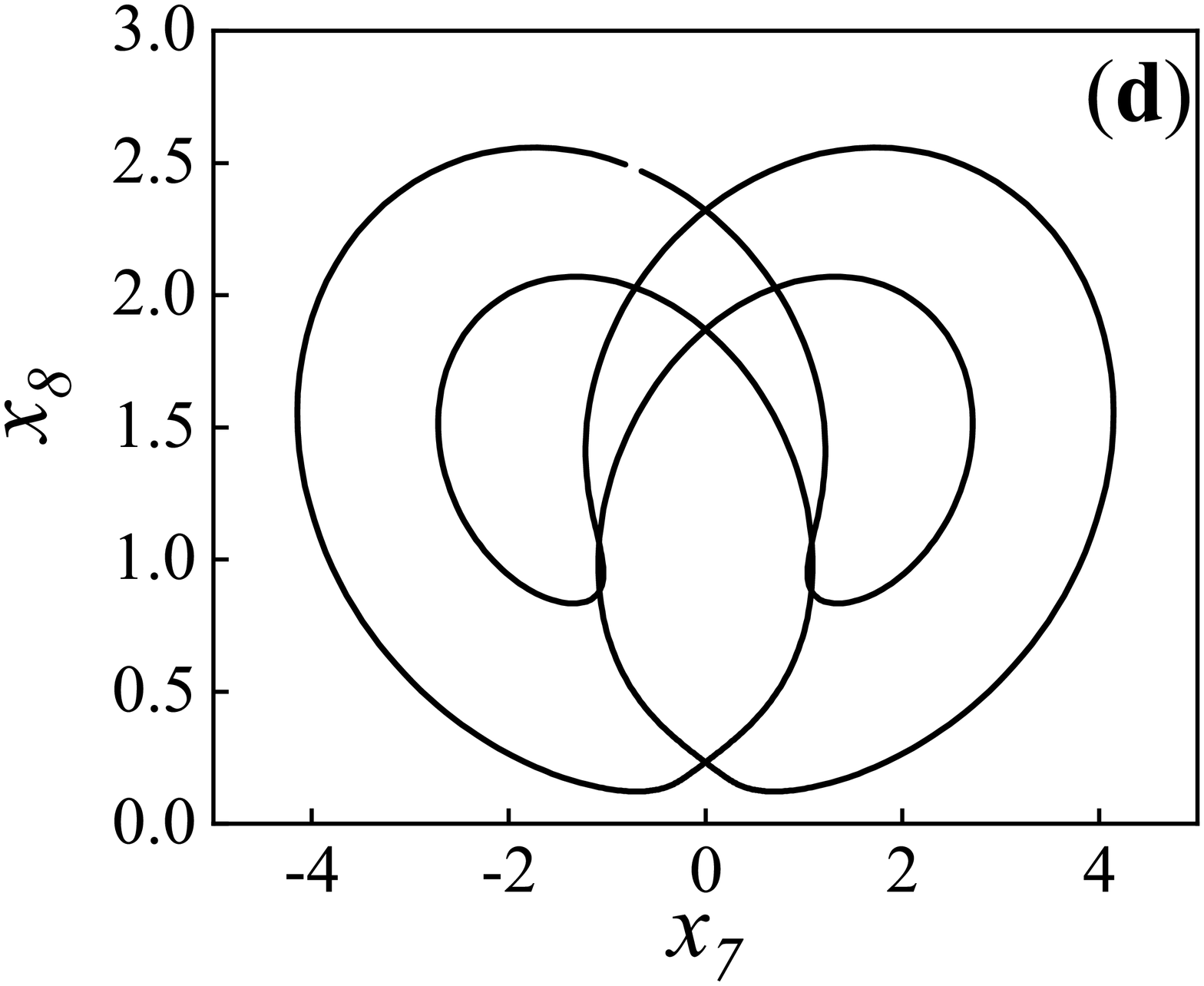}
  \caption{\label{L9}The initial guess and the detected periodic orbit of a reduced version Eq.~(\ref{ODE91}) of the Navier-Stokes equation. 
  (a) and (b) the projections of the initial guess loop in the $x_1-x_4$ and the $x_7-x_8$ plane, respectively. (c) and (d) for the periodic orbit obtained by the reduced variational method, with the parameter $\lambda=0.2216$ (indicating the period is 1.3924), $R_e=58$, the number of grid points $N=600$ and the dimension of the reduced coordinate frame $n=4$.}
\end{figure}
\begin{table}[]
\caption{\label{tab:ODE9}The Lyapunov exponents of the dynamical system described by Eq.~(\ref{ODE91}) for $Re=58$. }
\begin{ruledtabular}
\begin{tabular}{ccccccccc}
$\lambda_1$& $\lambda_2$ & $\lambda_3$ & $\lambda_4$ & $\lambda_5$ & $\lambda_6$ & $\lambda_7$ & $\lambda_8$& $\lambda_9$ \\
 0.61292 & 0.00143 &  -0.00069 & -1.81815 &   -3.74089 & -6.52364 &  -7.58614 & -8.61472& -14.33010 \\
\end{tabular}
\end{ruledtabular}
\end{table}

For the system described by Eq.~(\ref{ODE91}), we calculated numerically its Lyapunov exponents which are displayed in Table~\ref{tab:ODE9}. It can be seen that the first two values are greater than $0$. According to the previous discussion in Sec.\ref{sec:2_1}, the dimension of the local coordinate system is set to $n=4$ (the value of $\lambda_3$ and $\lambda_4$ are closest to $0$, which correspond to the orbit direction and the translation symmetry of Eq.~(\ref{ODE91})).
The results by the reduced method are displayed in Fig.~\ref{L9} in which (a) and (b) depict the projections of the initial guess loop constructed by the FFT, while (c) and (d) plot the projection of the final cycle. The initial guess loop does not seem to have any  symmetry, but the cycle is obviously symmetric.
As can be seen from diagrams (a) and (c), the initial loop roughly possesses the topology of the periodic orbit it is about to approach.
Therefore, even if the initial conditions are not very good, our method still drives  the guess loop to a periodic orbit.
\begin{figure}[]
  \includegraphics[width=8cm]{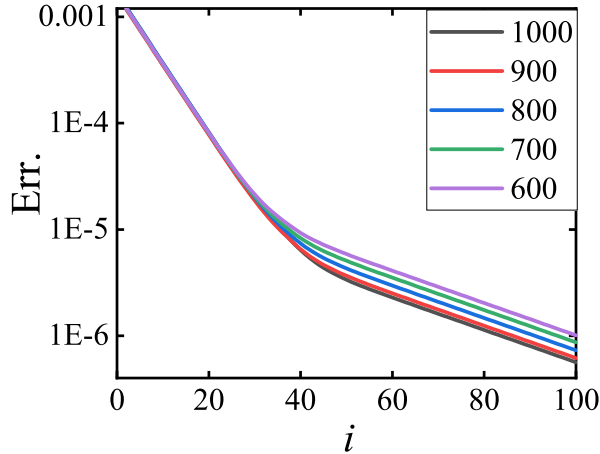}
  \caption{\label{Err9} The convergence of the reduced variational method for Eq.~(\ref{ODE91}) with
  $N=1000$, $900$, $800$, $700$, $600$, $Re=58$ and the dimension of reduced space is $n=4$}
\end{figure}
Here we further investigate the influence of the number of lattice points on the convergence. The same periodic orbit (as shown in  Fig.~\ref{L9}) is used as the target with different numbers of lattice points $N= 1000,~900,~800,~700,~600$ and the accuracy of the initial loop being kept the same. With the guess loop iterated for $100$ times, the convergence of our reduced method is shown in Fig.~\ref{Err9}. It is clear that in the first $40$ iterations, the convergence are similar among different representations, but becomes different in the ensuing iterations.
The larger the number of orbit points, the better the convergence. The fast convergence at the first stage stems from the initial exponential decay of the error in the projected subspace and thus is similar for different discretization. However, the relatively slow convergence at the second stage is enabled by  the modifications of the coordinate frames which proceed distinctively for  different $N$'s.

\subsection{The Coupled Lorenz Systems}
In the flowing we test our reduced method in case of the coupled Lorenz systems\cite{2007Frenzel} $\sum_i:\{\dot{x}_i(t),\dot{y}_i(t),\dot{z}_i(t)\}$, described by the following equations
\begin{align}\label{L99}
  \dot{x_i} =& \sigma(y_i-x_i), \nonumber\\
  \dot{y_i} =& x_i(\rho-z_i)-y_i+k_{i}y_{i+1}, ~~~~~~~~~~(i=1,2,3,...,\bar n)\\
  \dot{z_i} =& x_iy_i-\mu z_i ,\nonumber
\end{align}
where the parameters $\sigma=10$, $\rho=28$, $\mu=8/3$ with a periodic boundary condition: $y_{\bar{n}+1}=y_1$. Here, $\bar n$ is the number of the coupled lorenz systems and $k_i$ is the coupling strength. In the uncoupled case $k_i=0$, all the subsystem perform chaotic motions on the well-known Lorenz attractor. Linear coupling is imposed via the $y_{i+1}$ components of the adjacent subsystem, controlled by $k_i$. Here, we take the case of $\bar n=33$ as an example, where the total dimension of the whole system is $d=99$. In order to make the coupling of the system non-uniform, we set the coupling strength as $k_i= [0.5+0.5\frac{(i-1)}{(\bar n-1)})]k_0$ (with $k_0=0\sim 1$).

 \begin{table}[]
\caption{\label{tab:table1}The Lyapunov exponents of the $33$ coupled Lorentz systems with $k_0=1$. Here we only show the first 35 Lyapunov exponents.}
\begin{ruledtabular}
\begin{tabular}{cccccccccccc}
$\lambda_1$& $\lambda_2$ & $\lambda_3$ & $\lambda_4$ & $\lambda_5$ & $\lambda_6$ & $\lambda_7$ & $\lambda_8$& $\lambda_9$& $\lambda_{10}$ & $\lambda_{11}$& $\lambda_{12}$\\
1.3726 & 1.2824 &1.2119&  1.1894 & 1.1337 & 1.1154& 1.0835  & 1.0109 &0.9931 & 0.9299 &0.8913&0.8822 \\
\colrule
   $\lambda_{13}$ & $\lambda_{14}$ & $\lambda_{15}$ & $\lambda_{16}$ &$\lambda_{17}$& $\lambda_{18}$ & $\lambda_{19}$ & $\lambda_{20}$ & $\lambda_{21}$ & $\lambda_{22}$&$\lambda_{23}$ & $\lambda_{24}$ \\
   0.8017 &  0.8005 & 0.7710 & 0.6820 & 0.6550  &  0.6428 & 0.5747 & 0.5522 &0.5258 & 0.4586& 0.4470 & 0.3925 \\
\colrule
   $\lambda_{25}$& $\lambda_{26}$& $\lambda_{27}$& $\lambda_{28}$& $\lambda_{29}$& $\lambda_{30}$& $\lambda_{31}$& $\lambda_{32}$ & $\lambda_{33}$& $\lambda_{34}$& $\lambda_{35}$&...\\
 0.3609 &  0.3086& 0.2899& 0.2871 & 0.2331& 0.2098&0.1945& 0.1635 &  0.1161 &0.0176& 0.01184&... \\
\end{tabular}
\end{ruledtabular}
\end{table}
\begin{figure}[]
  \includegraphics[width=5cm]{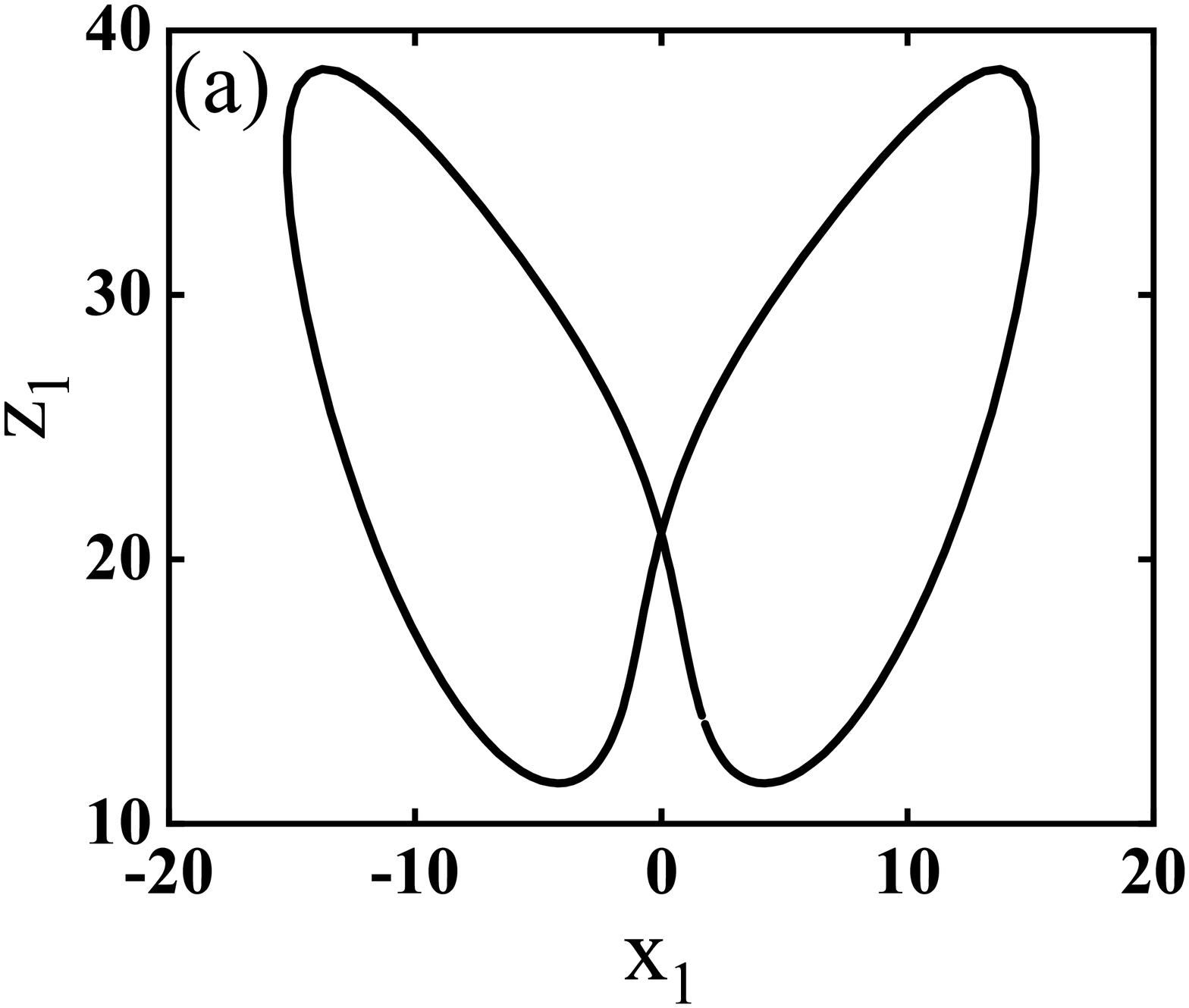}
  \includegraphics[width=5cm]{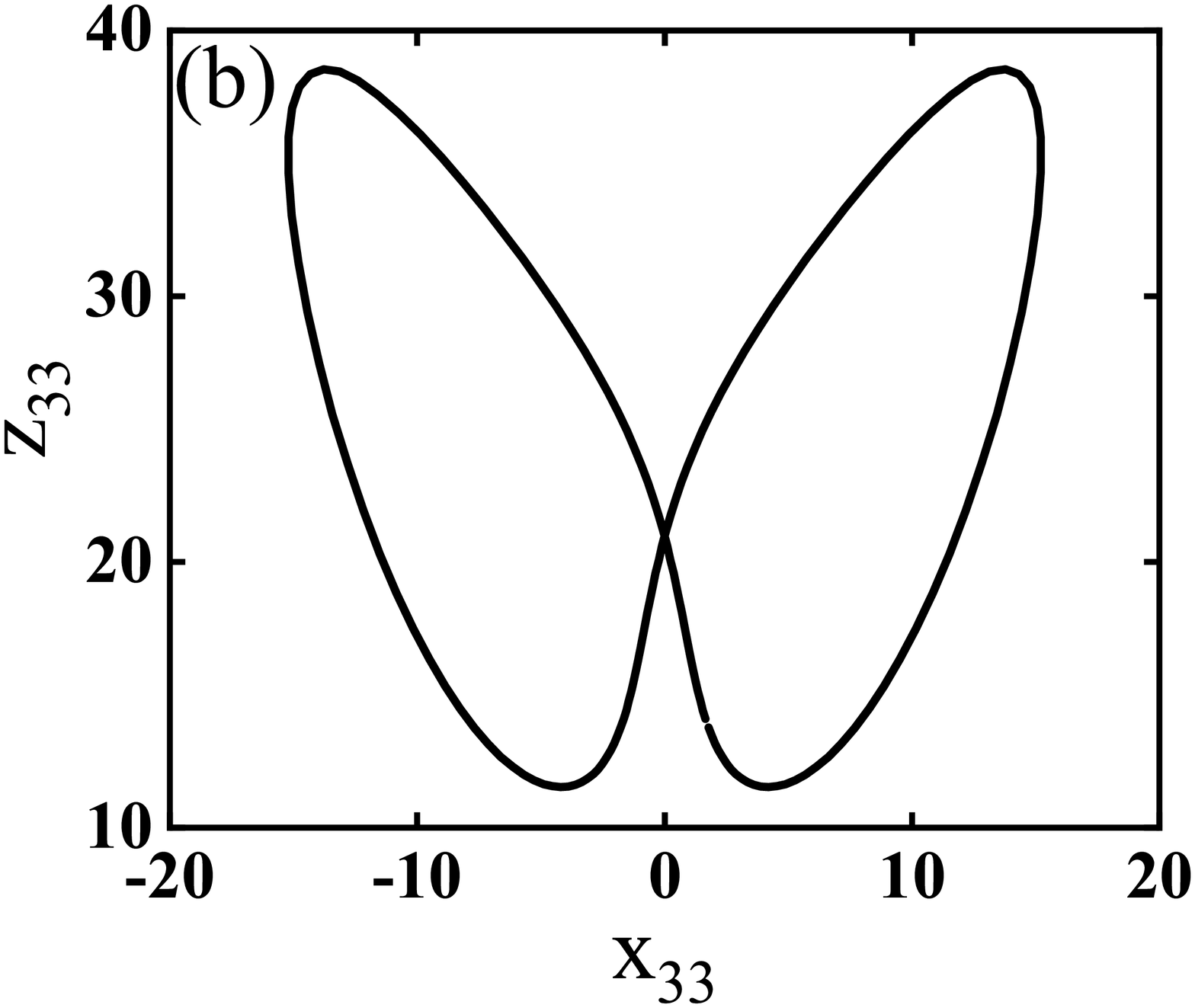}
  \includegraphics[width=5cm]{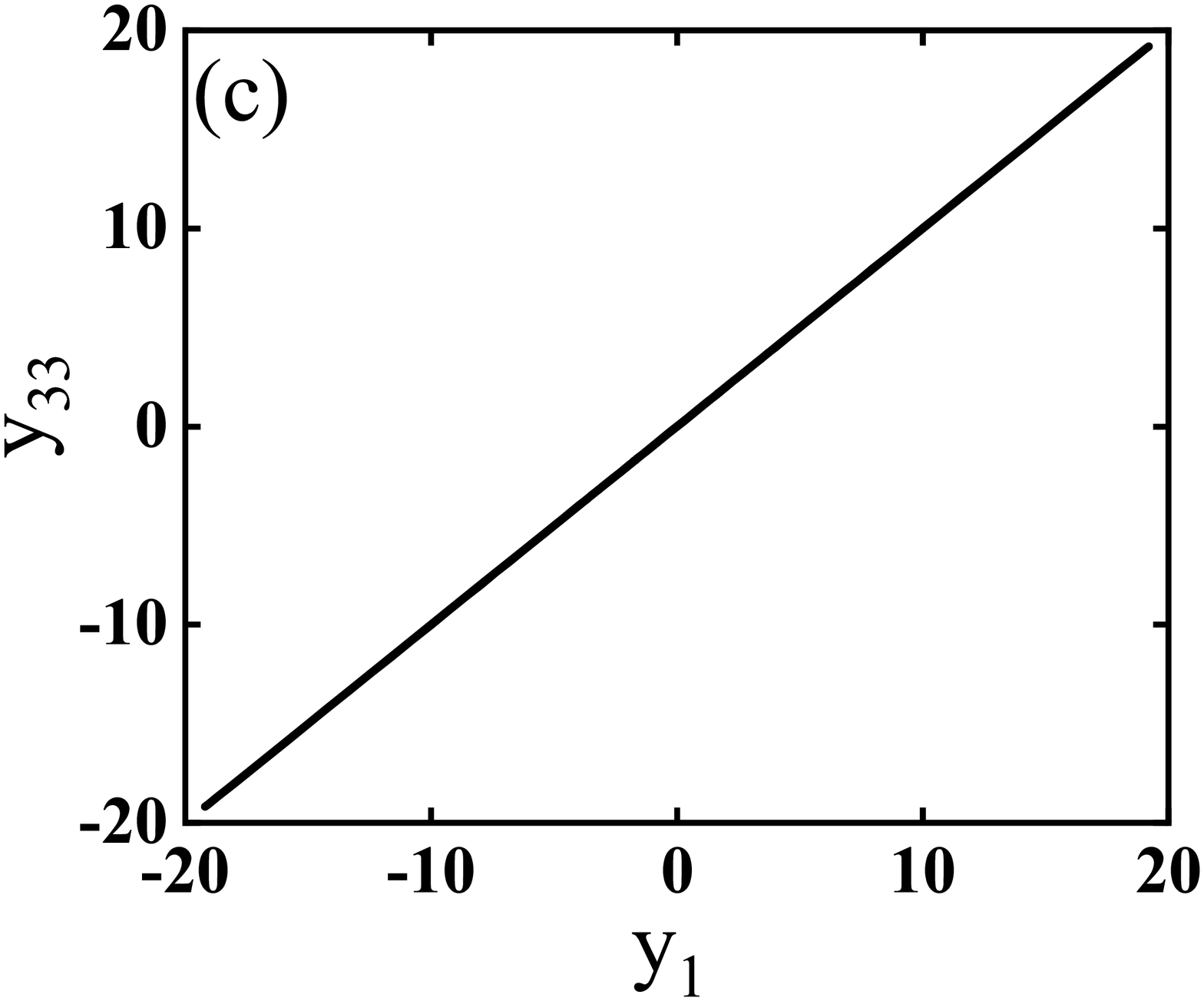}
  \includegraphics[width=5cm]{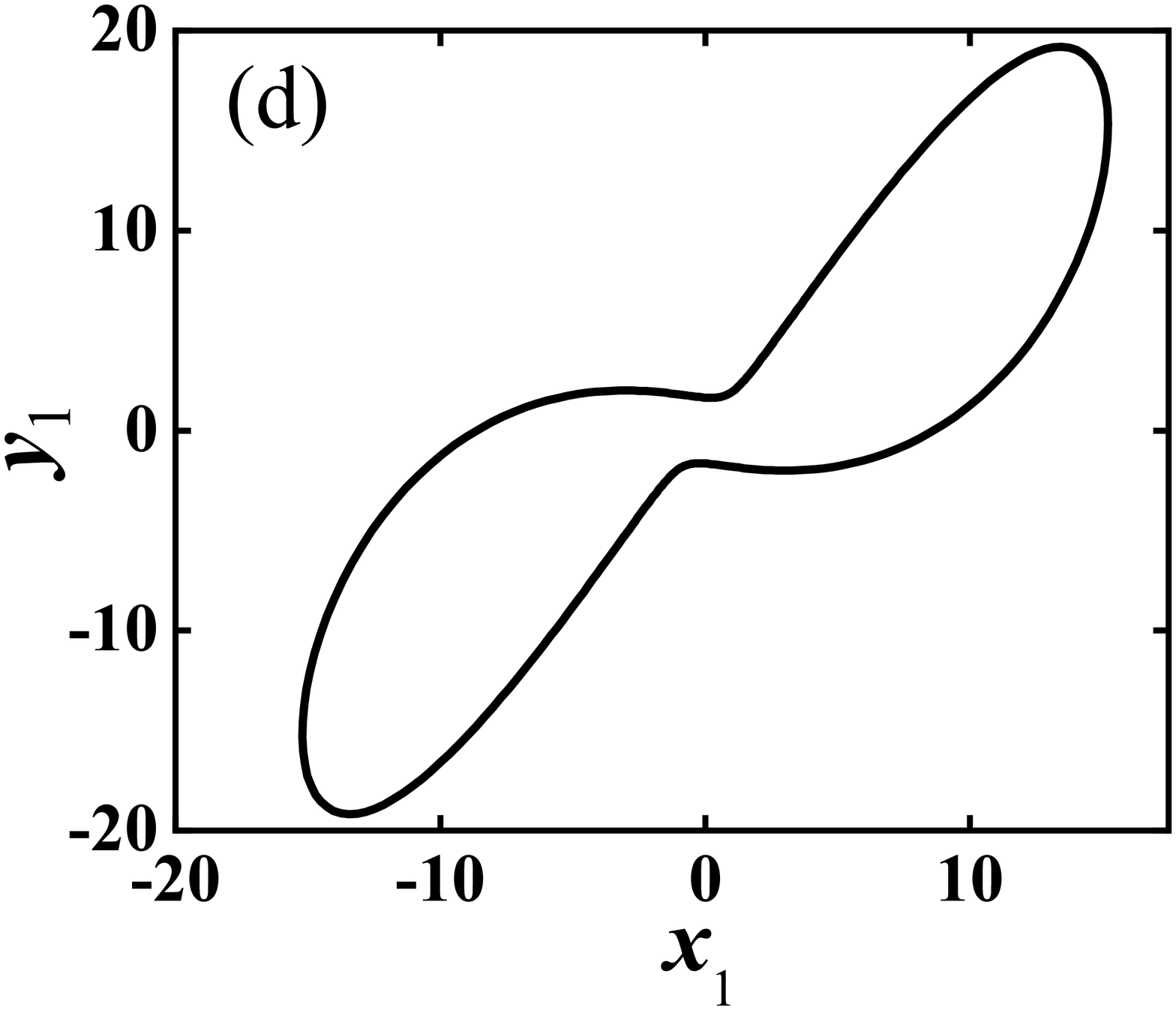}
  \includegraphics[width=5cm]{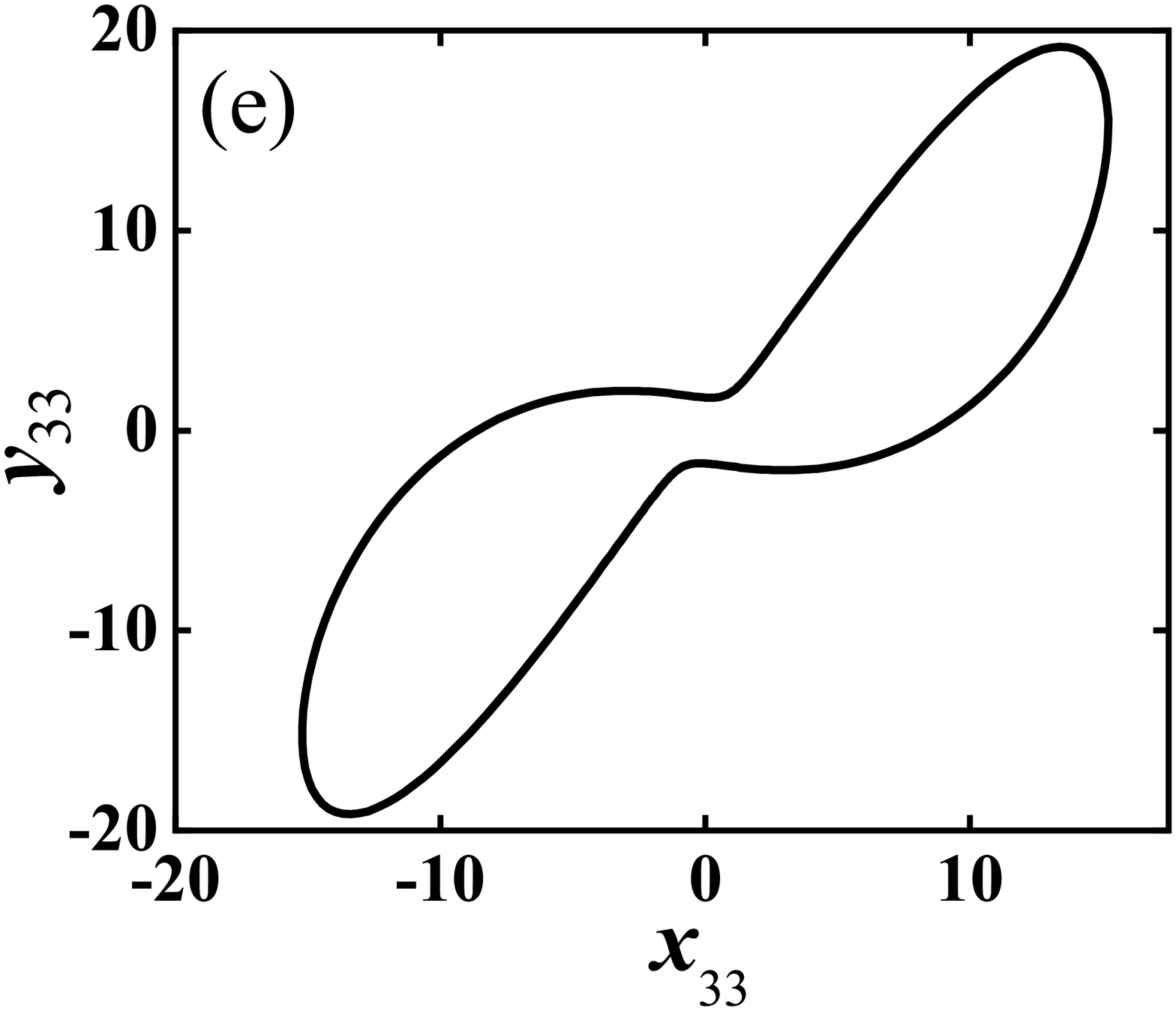}
  \includegraphics[width=5cm]{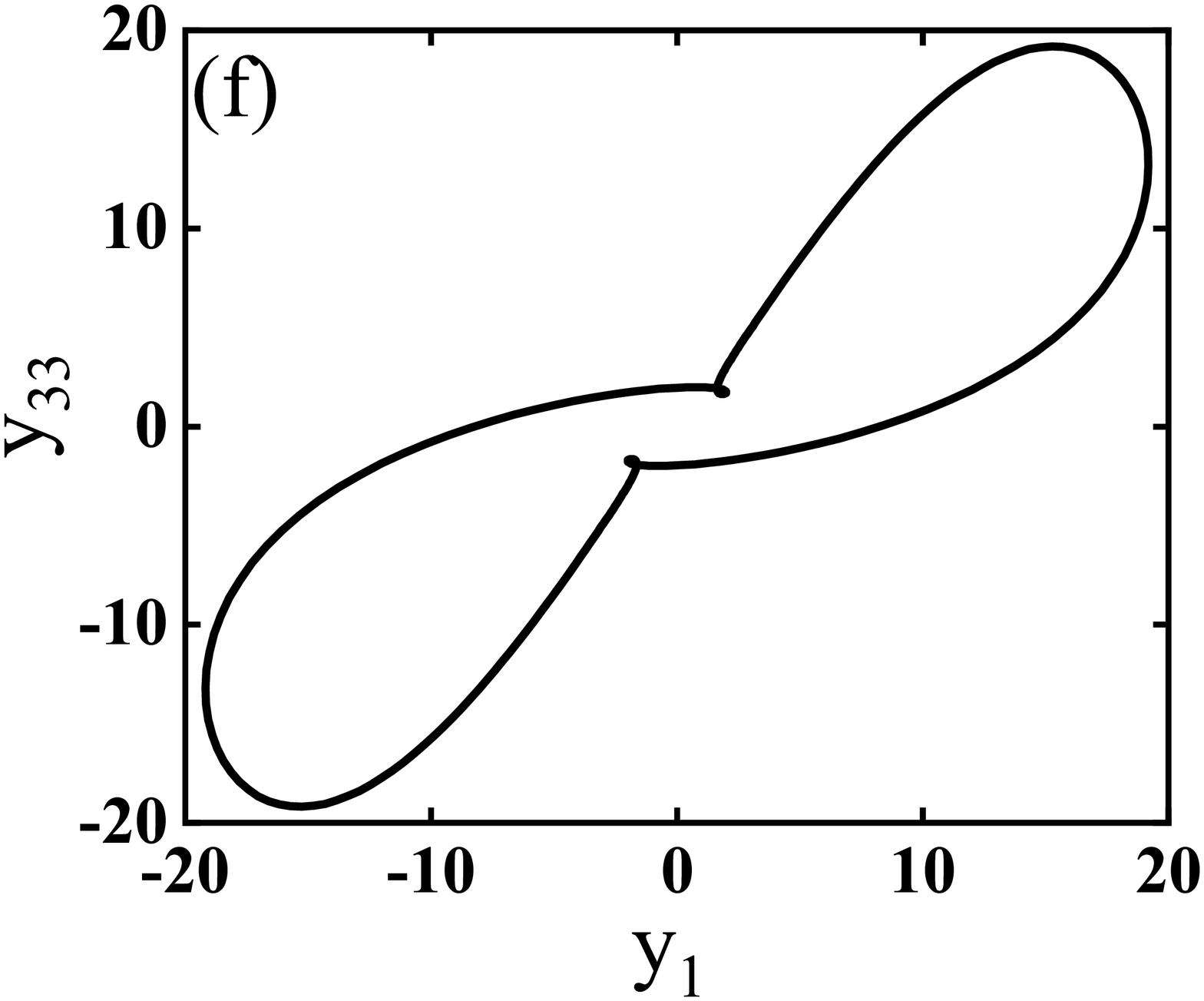}
  \caption{\label{L99_6} Periodic orbits of the coupled lorentz systems. (a) and (b), the periodic orbits for the first and the 33\emph{rd} Lorenz system with the coupling $k_i=0$ and $T=1.5997$;  (c) the linear relation of the variable $y_1$ and $y_{33}$ during the evolution; (d) and (e), the periodic orbit obtained by the reduced method with the coupling $k_0=1$, $T=1.5915$, with $N=600$ and $n=34$;  (f) the nonlinear dependence of $y_1$ on $y_{33}$.}
\end{figure}

For such a system, we may employ an alternative way of constructing an interesting initial loop. Since the system consists of $33$ coupled lorentz oscillators, we may use periodic orbits of the original lorentz system to construct the initial loop. For example, we take the same cycle for each lorenz system (as shown in Fig.\ref{L99_6} (a)$\sim$(c), the periodic orbit is symmetrical with respect to $x = 0$). When the coupling strength is $k_i=0$, the initial guess is already a true periodic orbit of the whole system.
Recursively, we search a nearby new cycle upon gradually increasing the coupling constant, starting with the already determined cycle at the previous coupling. This provides an alternative way to start the search and a family of cycles are obtained when $k_0$ goes from $0$ to $1$.
We have numerically calculated the Lyapunov exponents displayed in the Table \ref{tab:table1}, and found that the values of the first $33$ Lyapunov exponents are positive and away from zero corresponding to the main stretching directions. Such a result is also easy to understand, because there is an expanding direction in each Lorentz system, therefore the $33$ coupled Lorentz systems should have $33$ unstable directions as long as the coupling is small. According to the previous discussion in Sec.\ref{sec:2_1}, the dimension of the reduced space is set as $n=34$.

\begin{figure}[]
  \includegraphics[width=8cm]{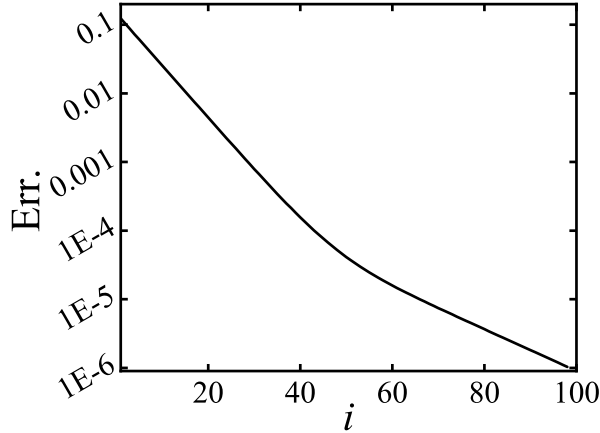}
  \caption{\label{Err99} The convergence of the reduced variational method for the coupled lorenz system with $k_0=1$ and the dimension of the reduced space is $n=34$}
\end{figure}

 The final result is shown in Figs.\ref{L99_6} (d) $\sim$ (f). By comparing the graphs in (d) and (e), we see that the periodic orbit of each individual Lorentz system does not seem to change much in the $x-y$ plane.
 However, the relation between corresponding variables in different subsystems exhibit nontrivial features (see Figs.~\ref{L99_6} (f)).
 In this example, from the construction of the initial loop to the search of periodic orbit, the whole process is completed in local subspaces without evolution in the full phase space, which certainly cuts down the computation complexity and provides a good example to deal with high-dimensional systems with similar spatial structures. As shown in Fig.\ref{Err99}, the algorithm converges faster initially and slows down when the guess loop is about to approach the true periodic orbit for the same reason given in the previous example. But on the whole, the convergence is good and remains exponential.

 In previous example, the unstable directions are quite few so that the dimension of the reduced local coordinate subspace is also low, the convergence may be expected. Nevertheless, in the current example, there are many unstable directions but the convergence is still good with a proper construction of the local subspaces. As a matter of fact, the local dimension can be further reduced even down to $n = 9$ and the method is still effective though with a slow convergence. How is it possible? During the evolution, all the local coordinate frames are essentially different. Each modification rotates these frames to new dimensions and the variational method will bring down errors in these directions. Finally, when all the directions are covered, a cycle may be identified. Nevertheless, the number of the included dimensions could not be too small to make the algorithm unstable. We still have no idea of what the necessary number of dimension is.

\subsection{The Kuramoto-Sivashinsky system}\label{KSE}
\begin{figure}[]
  \includegraphics[width=7cm,height=5.60cm]{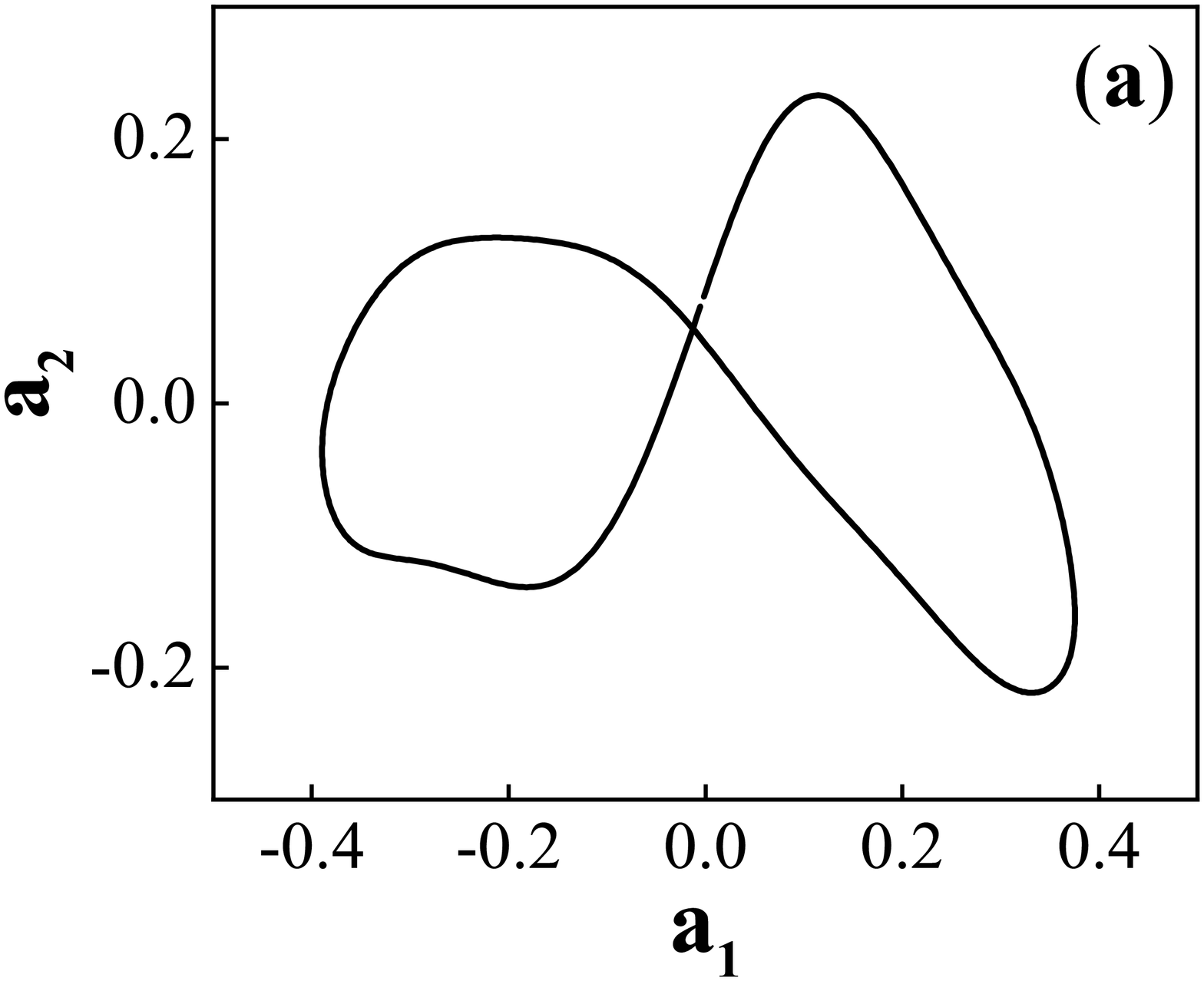}
  \includegraphics[width=7.1cm,height=5.80cm]{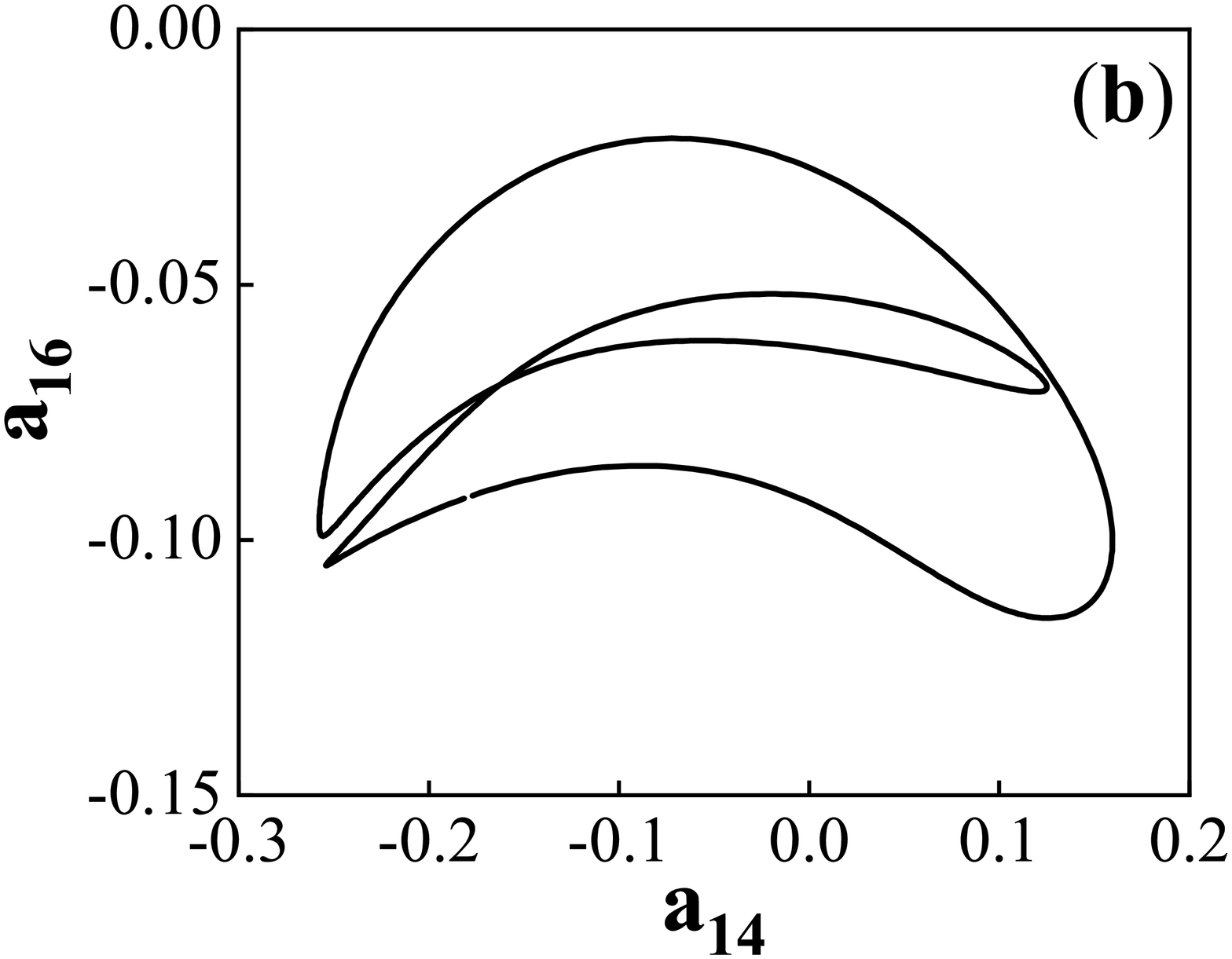}\\
  \includegraphics[width=7cm]{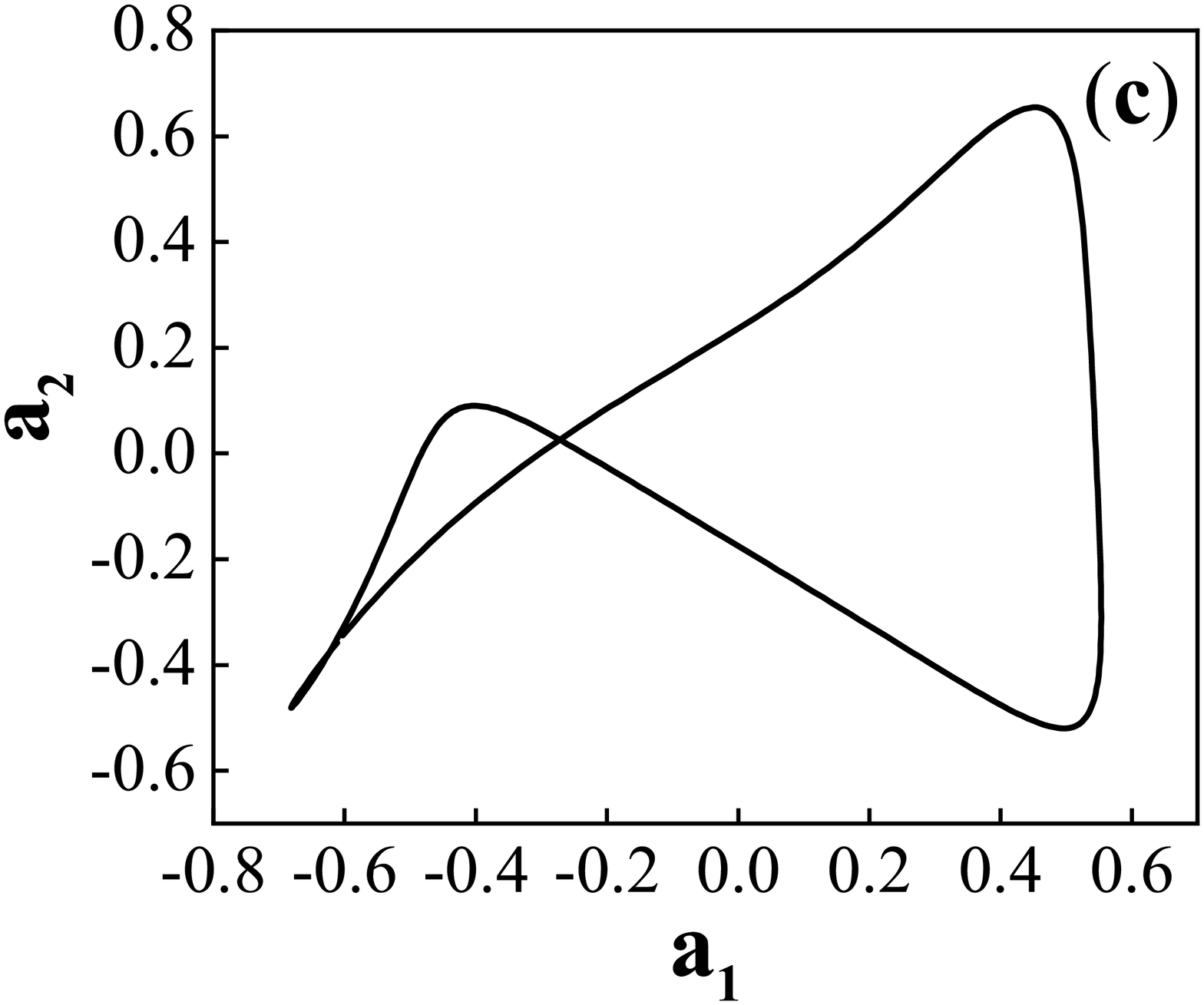}
  \includegraphics[width=7cm,,height=5.80cm]{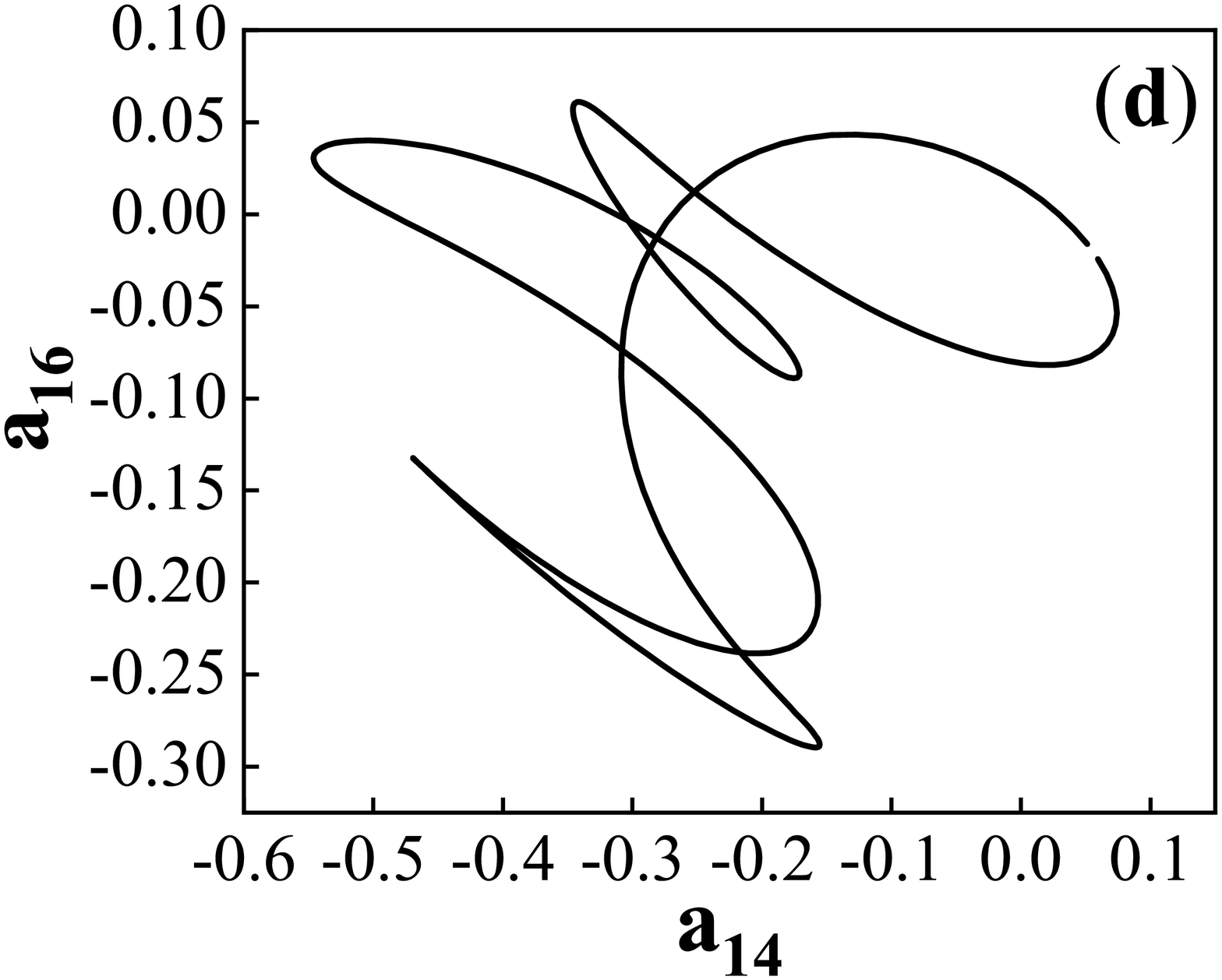}
  \caption{\label{ks16} The Kuramoto-Sivashinsky system in a spatiotemporally chaotic regime (viscosity parameter $\nu=0.015$, $d=16$ Fourier mode truncation). (a) and (b), the projection of the initial guess loop in the $x_{1}-x_{2}$ plane and the $x_{14}-x_{16}$ plane, respectively. (c) and (d), the projection of the resulting periodic orbit with $\lambda=0.0938$, $T=0.5892$, $N= 856$ and $n=5$.
  }
  \end{figure}
Here, we directly apply the current scheme to a nonlinear partial differential equation - the Kuramoto-Sivashinsky equation (KSe) to check the influcence of high dimensions and possible complications arising from possible stiffness due to the existence of inertial manifold. The KSe was independently proposed by Kuramoto and Tsuzuki in the study of phase turbulence in reaction-diffusion systems~\cite{KURAMOTO1976} and Sivashinsky in the study of flame combustion propagation models~\cite{Sivashinsky1977}. The KSe is an intrinsic phase equation describing the slow change of vibrations in a spatially extended system and also is an ideal model for studying high-dimensional systems. The KSe in one-dimensional space reads
\begin{equation}\label{KSe}
  u_t=(u^2)_x-u_{xx}-\nu u_{xxxx},
\end{equation}
where the subscript $x$ and $t$ represent the partial derivative of $u$ with respect to $x$ or $t$, with $t \geq 0$ and $x\in[0,2\pi]$.
The $\nu$ is a viscous damping parameter which is the dissipation rate of the system, and the first term on the right hand side of Eq.~(\ref{KSe}) is a nonlinear convection term, which induces interaction between different spatial modes and energy transfer from low wave-number modes to high ones. As $\nu$ decrease, the system undergoes a series of bifurcations, leading to increasingly turbulent dynamics.

If we choose to study only the odd solutions $u(t, -x)=-u(t, x)$ with the periodic boundary condition $u(t,x+2\pi)=u(t,x)$, the state variable $u(x,t)$ can be expanded in spatial Fourier series~\cite{2012Hopf},
\begin{equation}\label{Fourier}
  u(t,x)=i\sum\limits_{k=-\infty}^{\infty}a_k(t)e^{ikx},
\end{equation}
where $a_{-k}=-a_k\in\mathbb{R}$. In terms of the Fourier components,  Eq.~(\ref{KSe}) is rewritten as an infinite ladder of ODEs:
\begin{equation}\label{ODEFour}
  \dot{a}_k=(k^2-\nu k^4)a_k-k\sum\limits_{k=-\infty}^{\infty}a_ma_{k-m}.
\end{equation}
In numerical computation, a Galerkin truncation of the Fourier series has to be applied. As the amplitude of $a_k$ decreases with the increase of $k$ near the strange attractor, the high wavenumber modes in the asymptotic regime are negligible. Thus we can reduce the equation to a finite but large number of ODEs by the Galerkin truncation. In the current exploration, we work with two different truncations $d=16$ and $64$ in the following. However, in Eq.~(\ref{ODEFour}), the fourth power of $k$ arising from the dissipation could be quite large for large $k$, which makes the system stiff and brings trouble to numerical computation.

\begin{table}[]
\caption{\label{tab:Ks16}The Lyapunov exponents of The Kuramoto-Sivashinsky system with $d=16,~\nu=0.015$ and $d=64,~\nu=0.0049$. Here we only show the first 8 largest Lyapunov exponents. }
\begin{ruledtabular}

\begin{tabular}{ccccccccc}
\multirow{2}{*}{$d=16$} &$\lambda_1$& $\lambda_2$ & $\lambda_3$ & $\lambda_4$ & $\lambda_5$ & $\lambda_6$ & $\lambda_7$ & $\lambda_8$\\
  &12.8759 & 3.9500 & -0.0092 & -2.8395 &   -8.1252 &   -13.6478 &  -18.4199 & -23.5989 \\
 \colrule
 \multirow{2}{*}{$d=64$}&$\lambda_1$& $\lambda_2$ & $\lambda_3$ & $\lambda_4$ & $\lambda_5$ & $\lambda_6$ & $\lambda_7$ & $\lambda_8$\\
 &12.9183 & 7.8012 & 3.7726 & 0.3122 &   -0.0013 &   -3.2319 &  -8.1940 & -20.0568 \\
\end{tabular}
\end{ruledtabular}
\end{table}

The Lyapunov exponents of the system are computed and filled in Table.~\ref{tab:Ks16}.
We show the values of the first eight Lyapunov exponents at $d=16$ ($\nu=0.015$) and $64$ ($\nu=0.0049$), where the numbers of positive values are $2$ and $4$.
The exponents $\lambda_3=-0.0092$ and $\lambda_5=-0.0013$ correspond to the orbit direction of the system with $d=16$ and $d=64$, respectively.
According to the previous discussion in Sec.\ref{sec:2_1}, we select $n=5$ and $n=8$ directions to form the local coordinate frames along the loop, including the velocity direction.
\begin{figure}[]
  \includegraphics[width=7cm]{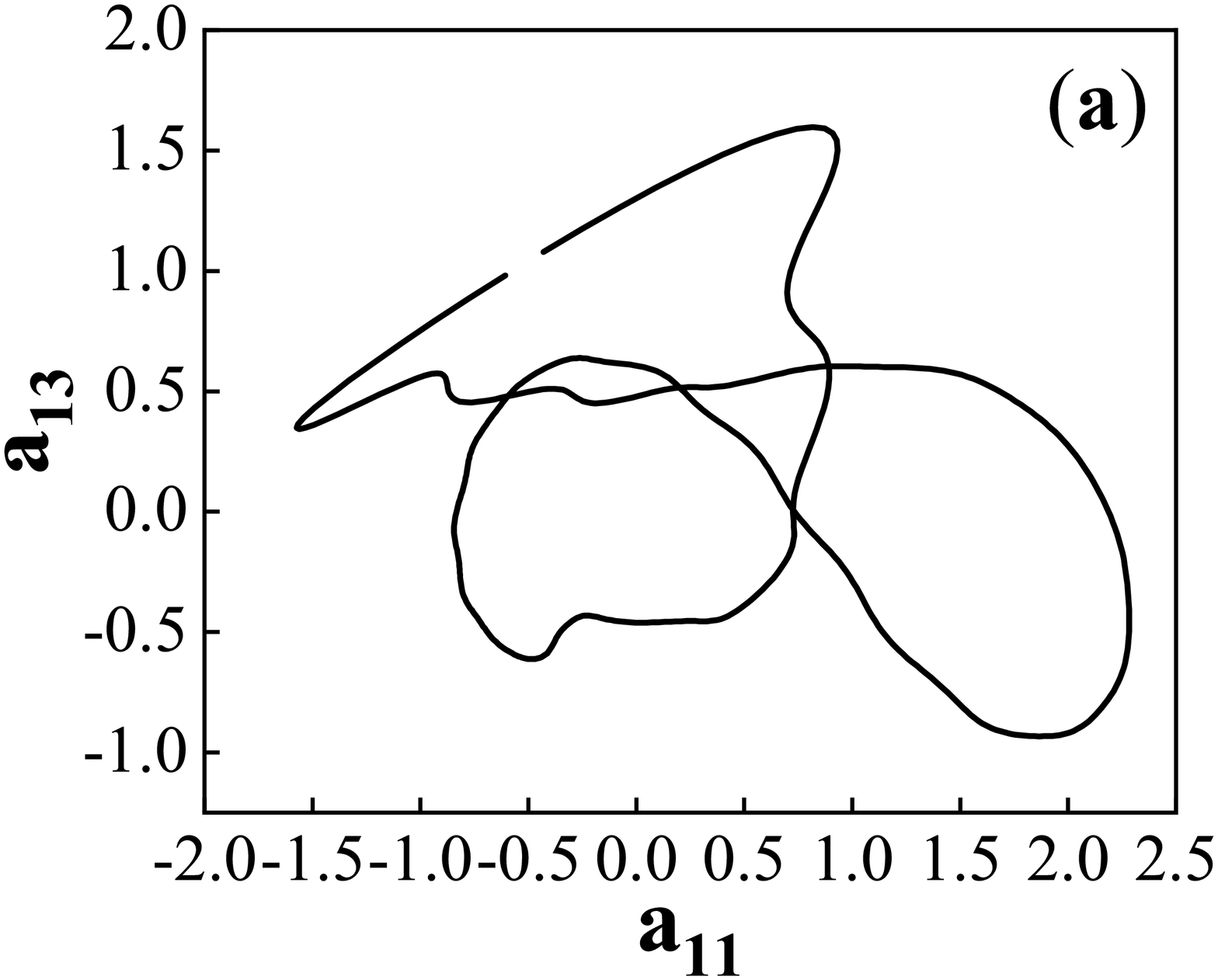}
  \includegraphics[width=7cm]{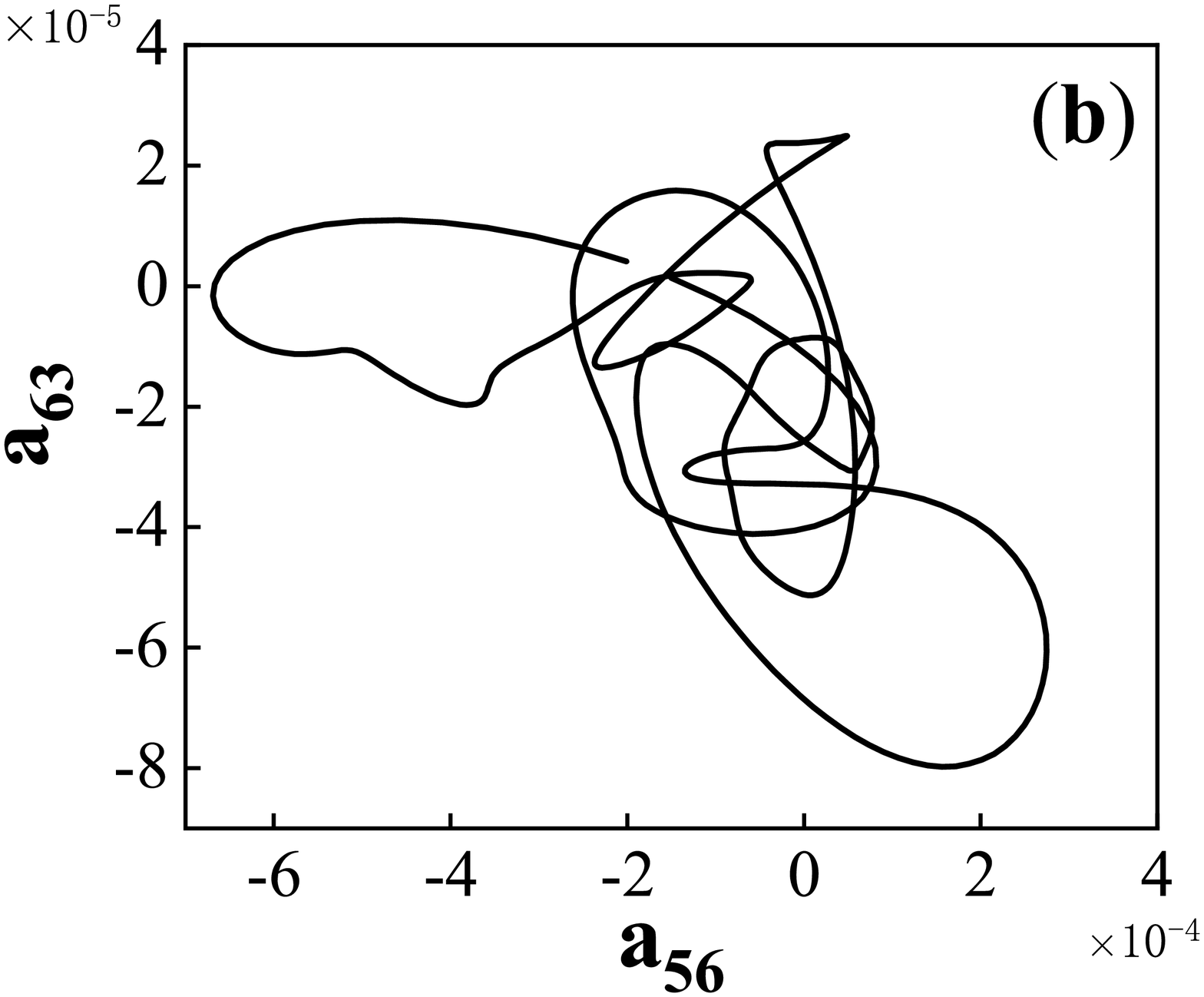}\\
  \includegraphics[width=7cm,height=5.80cm]{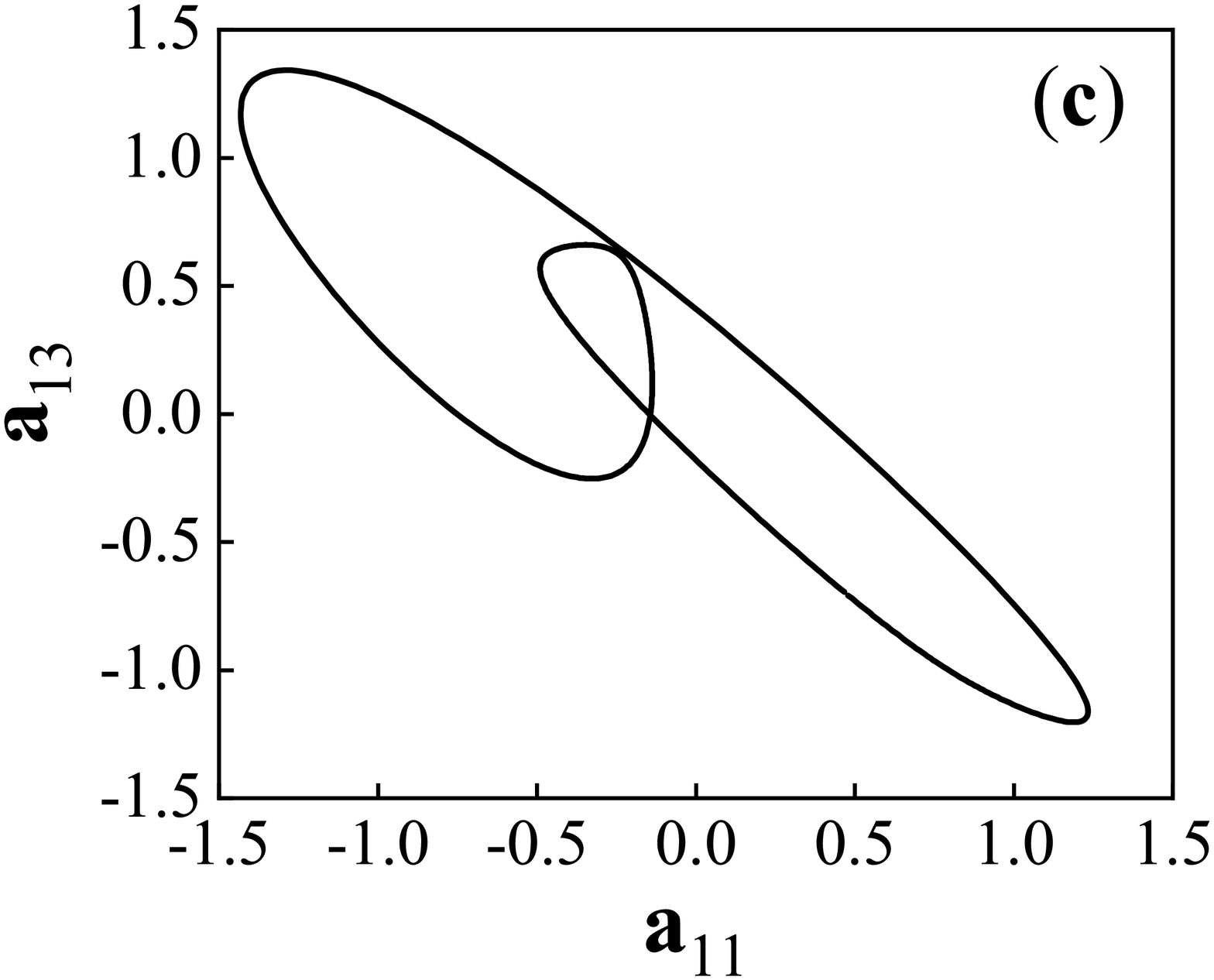}
  \includegraphics[width=7cm]{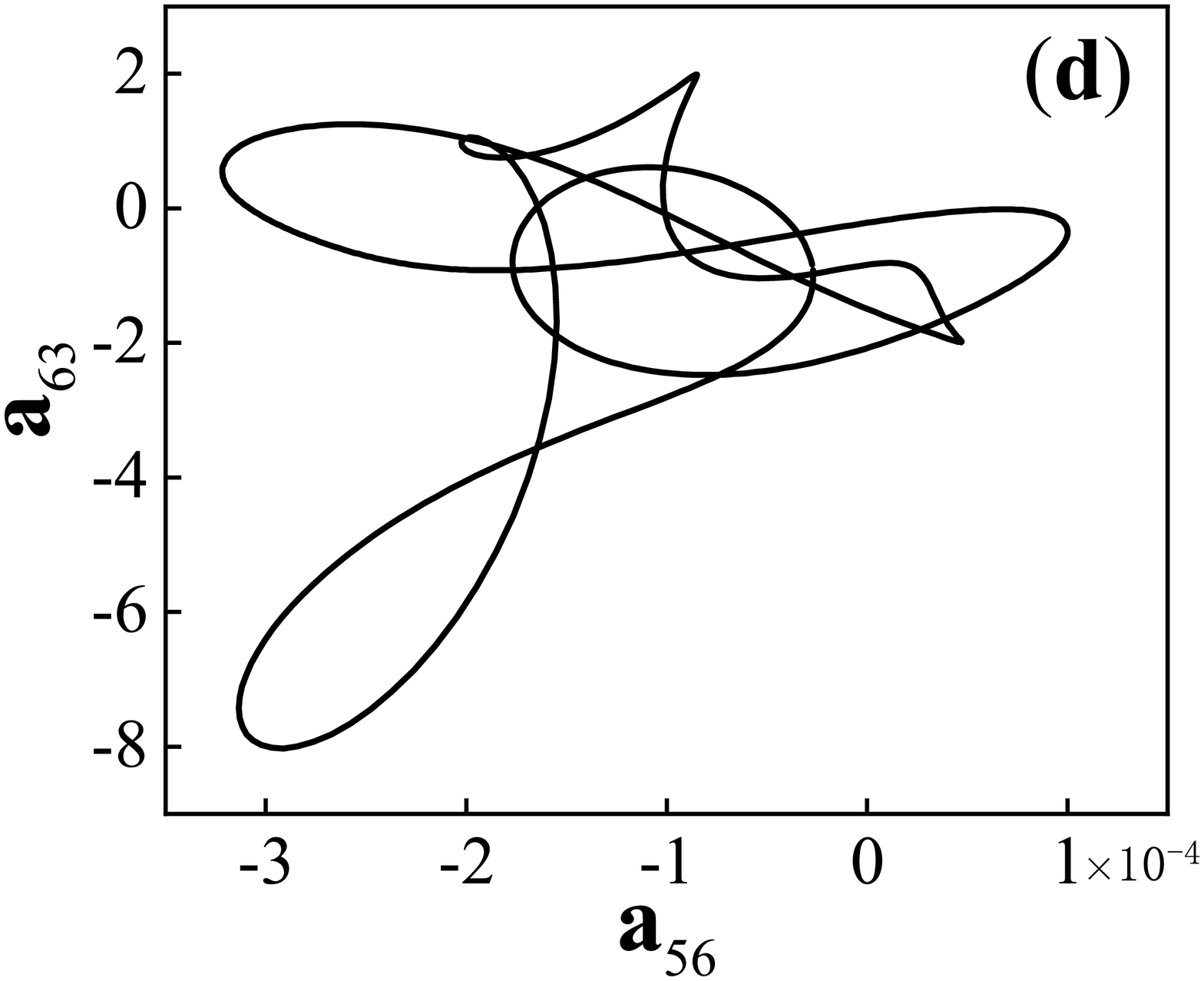}
  \caption{\label{ks64} The Kuramoto-Sivashinsky system in a spatiotemporally turbulent regime (viscosity parameter $\nu=0.0049$, $d=64$ Fourier mode truncation). (a) and (b), the projection of the initial guess loop on the $x_{11}-x_{13}$ plane and the $x_{56}-x_{63}$ plane, respectively. (c) and (d), the projection of resulting periodic orbit with $\lambda=0.0263$, $T=0.1649$, $N= 467$ (the number of grid points) and $n=8$ (The dimension of the reduced coordinate frame).
  }
\end{figure}
As shown in Figs.~\ref{ks16} and \ref{ks64}, the projections of the determined periodic orbits and the corresponding initial guess loops of with $d=16$ and $64$ are shown.
It is found that the amplitudes of high-wavenumber modes are much smaller than the low ones ($\bm a_{11}/\bm a_{63}\sim 10^4$).

Nevertheless, the high-wavenumber modes have intricate features, which makes them sensitive to the initial guess, especially when it is rough.
This is not surprising considering the stiffness problem mentioned after Eq.~(\ref{ODEFour}).
Therefore, extra care should be exercised during the modification of high-wavenumber modes
 the corrections of which could be multiplied by some
weight ranging from $1$ to $0.1$ to slow down the variation.
In addition, it is not difficult to find that the topological structures of the high-wavenumber modes change a great deal during loop evolution when the guess loop is not good, which prevents a smooth convergence and possibly induces numerical instability.
 With further increase of the dimensions of the system, this problem becomes prominent (such as in the case of $d=64$).
 \begin{figure}[]
  \includegraphics[width=8cm]{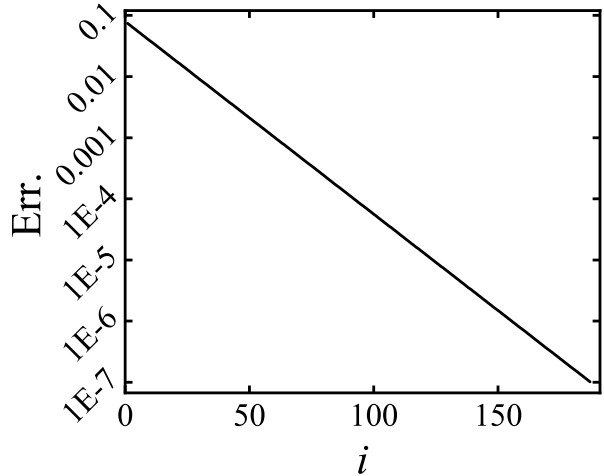}
  \caption{\label{Err64} The convergence of the reduced variational method for the Kuramoto-Sivashinsky system with $\nu_0=0.0049,~d=64$ Fourier mode truncation and the dimension of reduced space is $8$.}
\end{figure}
In order to promote the robustness of the reduced method the average position of the neighboring points on the loop is used as the new value at $s_i$ (namely, $\bm X(s_i)\longrightarrow[\bm X(s_{i})+\bm X(s_{i+1})]/2$) after each correction. Despite all the details dealing with a bad initial guess, in the Kuramoto-Sivashinsky system with $d = 64$ Fourier mode truncation, for a subspace with dimension $n=8$, our reduced method has good convergence. As shown in Fig.~\ref{Err64}, when the initial guess loop is close to a periodic orbit (entering the linearization neighborhood of the orbit), the error function Err. will decreases exponentially.

\begin{figure}[]
  \includegraphics[width=7cm]{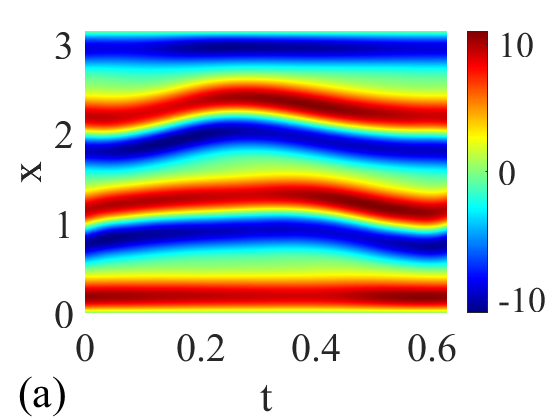}
  \includegraphics[width=7cm]{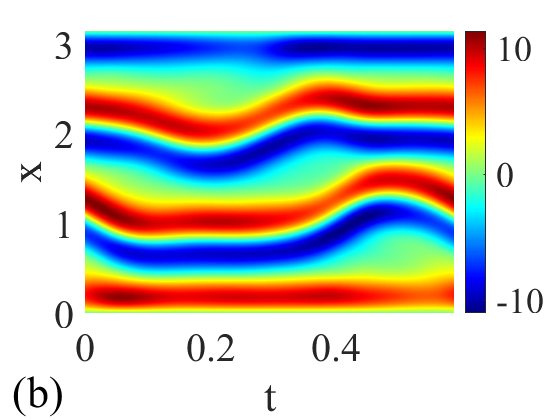}
  \includegraphics[width=7cm]{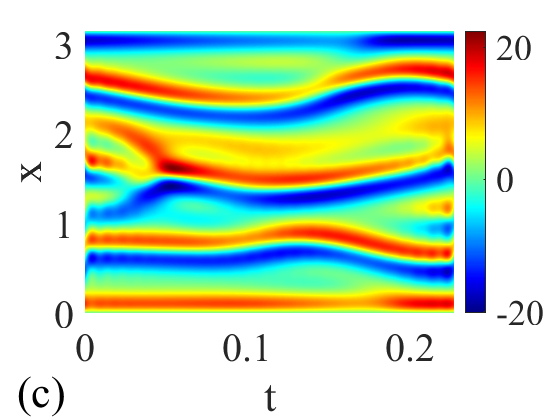}
  \includegraphics[width=7cm]{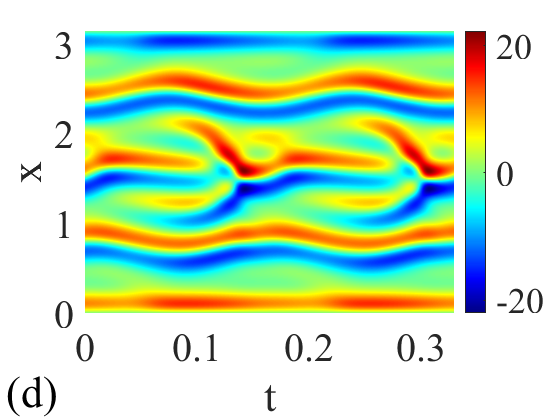}
  \caption{\label{ks64Fig} Level plot of the space-time evolution u(x,t) in the (t, x) plan. (a) and (c), the initial guesses in Fig.~\ref{ks16} and Fig.~\ref{ks64} respectively. (b) and (d), the periodic orbits depicted in  Fig.~\ref{ks16} (with the cycle period $T=0.5892$ ) and Fig.~\ref{ks64} (a repeat followed after the cycle period $T= 0.1649$ ).}
\end{figure}
The space-time evolutions of $u(x,t)$, the unstable  spatiotemporally solutions corresponding to the orbits in Fig.~\ref{ks16} and Fig.~\ref{ks64}, are plotted in Fig.~\ref{ks64Fig}. As $u(x,t)$ is antisymmetric in $[-\pi,\pi]$, it suffices to display the solutions in the $[0,\pi]$ interval.
There is little difference between the spatiotemporal profiles in Fig.~\ref{ks64Fig} (a) and (b), which indicates that the guess loop is close to the periodic orbit. While big difference between (c) and (d) indicates that the initial guess is bad. A comparison of (b) and (d) reveals that the structure displayed in (d) has more features than that in (b) which indicates that the higher the dimension of the attractor, the finer the space-time evolution of the $u(x,t)$.
\begin{figure}[]
  \includegraphics[width=7cm]{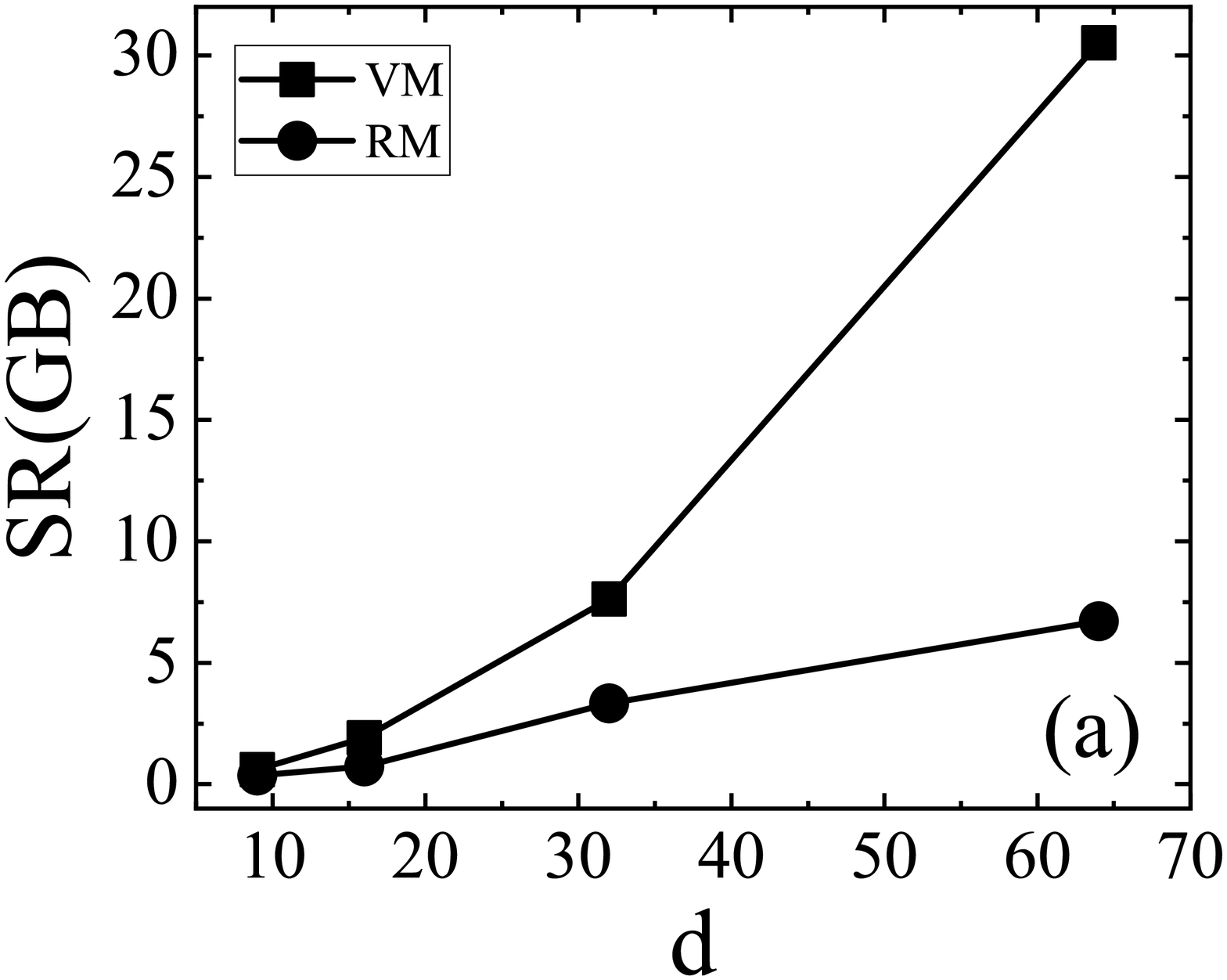}
  \includegraphics[width=7cm,height=5.80cm]{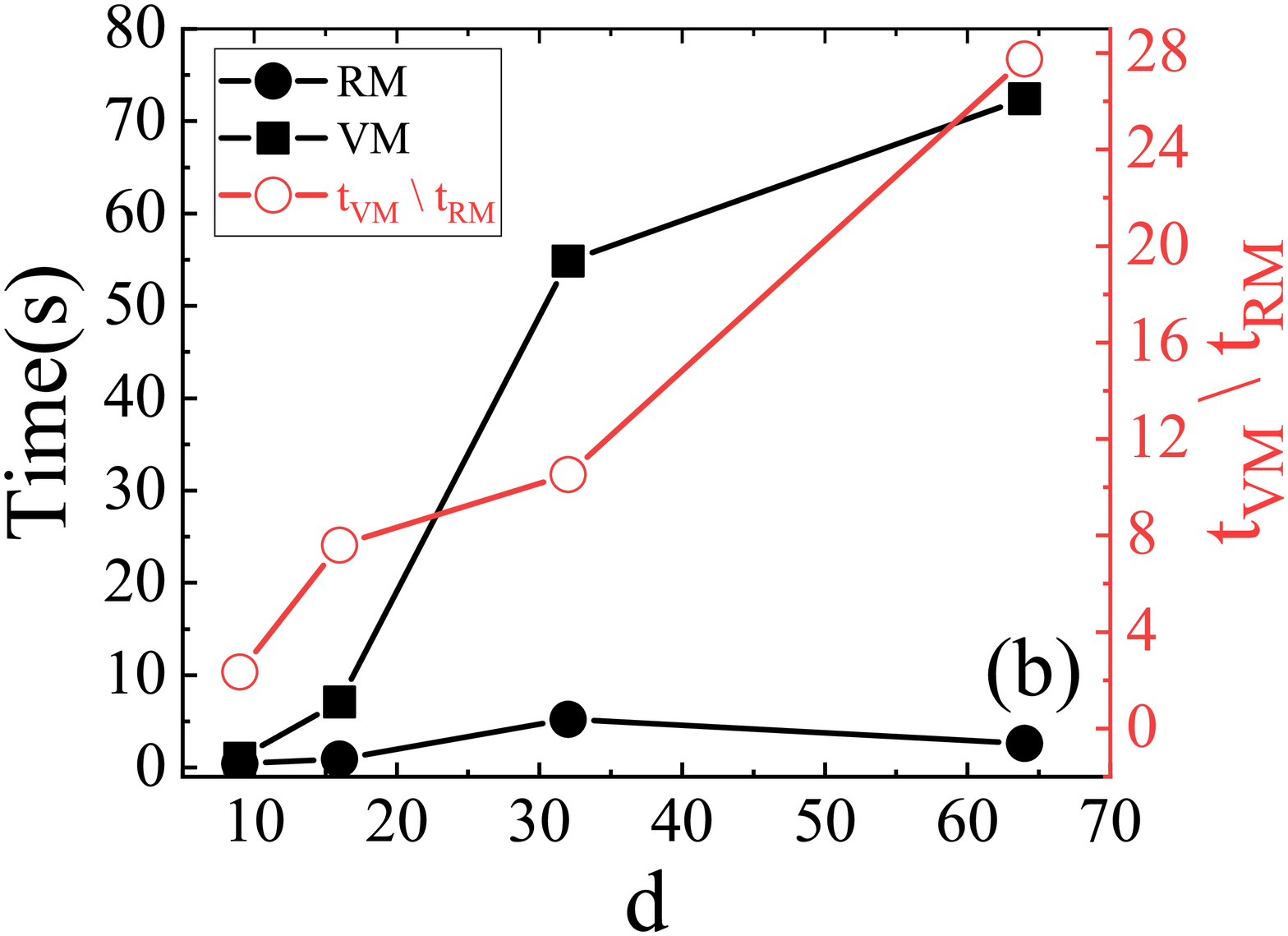}
  \caption{\label{RT} The estimated storage and time cost required by different variational approaches. (a) the storage resource (SR) required by the original variational method (VM, black-quare curve) and  by the reduced method (RM, black-dot curve). The number of lattice points is $10^3$. (b) the estimated time required by the RM (black-dot curve) and the VM (black-quare curve). The red-circle marks the ratio ($t_{VM}/t_{RM}$). The number of grid points is $746,~856,~856,$ and $467$ for systems with dimensions $d=9,~16,~32$ and $64$. It should be noted here that the system corresponding to $d=9$ is the previous nine-mode Lorentz system.
  }
\end{figure}

Often than not, the original variational method (VM) consumes a lot of storage resource (SR) and computer time which 
will become more and more striking with the increase of system dimension, but the reduced variational method (RM) introduced in the current manuscript partly solves the problem.
As shown in Fig.~\ref{RT}, we compared the required storage and computing time for the VM and the RM.
In Fig.~\ref{RT}~(a), the black-quare curve depicts the storage resource required by the VM, which increases exponentially with the increase of system dimension $d$, while the black-dot curve for the RM grows only linearly. The difference between them becomes more and more pronounced with the increase of system dimension $d$ (the ratio of storage resources required is $SR_{VM}/SR_{RM}\sim 6$ for $d=64$).
In Fig.~\ref{RT}~(b), the black-square curve is the time required by the VM, and the black-dot curve by the RM.
Because for different system dimensions, the number of lattice points is different, it is difficult to compare directly. A quantity $t_{VM}/t_{RM}$ (that is, the ratio of the time $t_{VM}$ required by VM to the time $t_{RM}$ required by RM)is defined and plotted with the red-circle curve in Fig.~\ref{RT}~(b). It is clear that with the increase of system dimension, the ratio $t_{VM}/t_{RM}$ becomes larger and larger, and reaches $t_{VM}/t_{RM}\sim 28$ at $d=64$. Certainly, when the dimension of the system is not high, the advantage of the RM is not so prominent.
Henceforth, the caveat that lies in the variational method is essentially filled in the reduced scheme and we expect its fruitful application to high dimensional systems.

\section{SUMMARY}\label{summary}
The variational method proposed in Ref.~\cite{Lan2004} is very robust numerically and hence widely used when the system dimension is not high.
However, in application to complex systems with high dimensions, its bottleneck of computation complexity is becoming prominent. That is, in solving the associated matrix equation, the demand for storage and computing time increases sharply with the increase of the phase space dimension. In order to break the bottleneck, we proposed a reduction scheme in the current work. The main idea is to reduce the high-dimensional coordinate frame at each loop point to a low-dimensional one based on the existence of inertial manifold widely observed in various spatially extended systems~\cite{Temam1988}. When the guess loop resides in the neighborhood of a periodic orbit, to enable the convergence, the unstable directions need to be carefully controlled, while the stable directions drag the loop to the periodic orbit during forward evolution.
Based this consideration, we give a reduced version of the variational method to take care of all the unstable directions but implement the natural evolution in the stable ones. In addition, a pseudo-evolution equation along the loop and its improved version are proposed to evolve the local coordinate systems. An alternating evolution of the loop and the local coordinate frames is practiced and observed to lead to robust convergence.

 In all examples, a rough criterion for determining the dimension $n$ of the reduced space is verified and should be effective for general systems that $n$ is greater than the number of positive Lyapunov exponents of the system.
 The velocity direction needs to be included in the construction of the local coordinate system, because in the early iteration of the guess loop, together with the expanding directions it enables a fast convergence.
 However, other directions may take the lead in subsequent iterations, which is also the reason why there is an inflection point on the convergence curve.
 Therefore, in order to make the algorithm stable and converging well, the alternation between the loop evolution ($10\sim20$ iterations) and the modification of local coordinates (about $5\sim15$ steps) is carried out and found to be effective.
 Away from the periodic orbit, the reduced scheme could become unstable just like the original variational method.
But through an averaging process, the stability is restored quickly as discussed in Section \ref{KSE} in its application to the $64$-dimensional Kuramoto-Sivashinsky system. When the guess loop enters the linearization neighborhood of a periodic orbits, the algorithm shows very good exponential convergence.

In the new reduced scheme, the evolution of local coordinate systems is based on pseudo-evolution. If the initial loop is poor, it is hard to pin down the stretching directions so that the convergence is quite slow, which makes the algorithm inefficient. Better schemes should be designed to utilize and benefit from possible low-dimensional structures.
Nevertheless, the reduced scheme shows superior stability and efficiency for searching periodic orbits in high-dimensional systems, and seems to provide a new tool to explore complex system dynamics dominated by recurring patterns.
\acknowledgments
This work was supported by the National Natural Science Foundation of China under Grants No. 11775035, and also by the Fundamental Research Funds for the Central Universities with Contract No.2019XD-A10.

\bibliography{Ref1}

\end{document}